\documentclass[sageh,times,Afour]{sagej} 

\usepackage[hyphens]{url}
\usepackage{moreverb}
\usepackage{amsmath,amssymb,amsfonts}
\usepackage{mathtools}
\usepackage{algorithmic}
\usepackage{textcomp}
\usepackage{xcolor}
\usepackage{flushend}

\usepackage{booktabs} \usepackage{etoc} \usepackage{multirow}
\usepackage{makecell}
\usepackage{enumitem}

\usepackage{graphicx} \usepackage[export]{adjustbox}
\usepackage{subfig}
\usepackage{epstopdf}
\DeclareGraphicsExtensions{.eps.pdf,.png,.jpg,.tif,.tiff,.ps}
\graphicspath{{figs/}}

\setcitestyle{authoryear}

\usepackage{booktabs} \usepackage{subfig}
\usepackage{enumitem}
\usepackage{multirow}
\usepackage[colorlinks,bookmarksopen,bookmarksnumbered,citecolor=red,urlcolor=red]{hyperref}
\usepackage[nameinlink,capitalize]{cleveref}

\usepackage{soul}
\usepackage[colorinlistoftodos,prependcaption,textsize=tiny,textwidth=10mm]{todonotes} 

\definecolor{navy}{rgb}{0.1, 0.1, 0.8}
\definecolor{gray}{rgb}{0.6, 0.6, 0.6}
\definecolor{myblue}{rgb}{.8, .8, 1}
\definecolor{olive}{rgb}{0.1, 0.5, 0.1}

\usepackage[normalem]{ulem}
 \newcommand\reduline{\bgroup\markoverwith{\textcolor{red}{\rule[0.5ex]{2pt}{0.4pt}}}\ULon}

\newcommand\BibTeX{{\rmfamily B\kern-.05em \textsc{i\kern-.025em b}\kern-.08em
T\kern-.1667em\lower.7ex\hbox{E}\kern-.125emX}}

\def\volumeyear{2021}
\def\volumenumber{XX}
\def\issuenumber{X}
\def\journalname{Journalism}
\def\DOI{\href{https://doi.org/10.1177/1464884921996286}{10.1177/1464884921996286}}

\begin{document}
\etocdepthtag.toc{mtchapter}

\runninghead{Dawson et al.}

\newcommand{\titlename}{Layoffs, Inequity and COVID-19: A Longitudinal Study of the Journalism Jobs Crisis in Australia from 2012 to 2020}
\title{\titlename}

\author{Nik Dawson\affilnum{1}\affilnum{2}, Sacha Molitorisz\affilnum{3}, Marian-Andrei Rizoiu\affilnum{4} and Peter Fray\affilnum{5}}

\affiliation{\affilnum{1}Centre for Artificial Intelligence, University of Technology Sydney\\
\affilnum{2}OECD Future of Work Research Fellow\\
\affilnum{3}Centre for Media Transition, University of Technology Sydney\\
\affilnum{4}Data Science Institute, University of Technology Sydney\\
\affilnum{5}Private Media
}

\corrauth{Nik Dawson 
}

\email{nikolasjdawson@gmail.com}

\begin{abstract}
In Australia and beyond, journalism is reportedly an industry in crisis, a crisis exacerbated by COVID-19. However, the evidence revealing the crisis is often anecdotal or limited in scope. In this unprecedented longitudinal research, we draw on data from the Australian journalism jobs market from January 2012 until March 2020. Using Data Science and Machine Learning techniques, we analyse two distinct data sets: job advertisements (ads) data comprising 3,698 journalist job ads from a corpus of over 8 million Australian job ads; and official employment data from the Australian Bureau of Statistics. Having matched and analysed both sources, we address both the demand for and supply of journalists in Australia over this critical period.
The data show that the crisis is real, but there are also surprises. Counter-intuitively, the number of journalism job ads in Australia rose from 2012 until 2016, before falling into decline. Less surprisingly, for the entire period studied the figures reveal extreme volatility, characterised by large and erratic fluctuations. The data also clearly show that COVID-19 has significantly worsened the crisis.
We then tease out more granular findings, including: that there are now more women than men journalists in Australia, but that gender inequity is worsening, with women journalists getting younger and worse-paid just as men journalists are, on average, getting older and better-paid; that, despite the crisis besetting the industry, the demand for journalism skills has increased; and that, perhaps concerningly, the skills sought by journalism job ads increasingly include social media and generalist communications.
\end{abstract}


\maketitle
\section{Introduction}

Globally, the news about the news is not good. 
This was true before 2020, but COVID-19 has only made matters worse. 
Take Australia.
In March 2020, newswire service the Australian Associated Press announced it would be shutting down its operations after 85 years~\mbox{\citep{Samios2020}}.
In June, a last-minute consortium of investors and philanthropists saved the day -- but salvation was merely partial, with only 85 of the company's 180 journalists, photographers and other staff retained~\citep{Wahlquist2020}. 
Meanwhile, News Corp has been closing scores of regional titles (see below). 
In the US, the news about the news is just as bad, if not worse. 
In February, the country's No. 2 newspaper chain (McClatchy) declared bankruptcy~\citep{Benton2020}. 
Amid widespread pay cuts, furloughs and layoffs, US newsrooms reportedly shed more than 11,000 jobs in the first half of 2020~\citep{Willens2020}.

Even before COVID-19, digital technology upended journalism's advertising-driven business model~\citep{ACCC2019}. 
As the Nieman Lab notes: 
\begin{quotation}
The Internet has brought forth an unprecedented flowering of news and information. But it has also destabilised the old business models that have supported quality journalism for decades. Good journalists across the country are losing their jobs or adjusting to a radically new news environment online~\citep{Nieman-Lab}.\end{quotation} 
But is journalism \textit{in crisis}? A wealth of research in Australia, the US and comparable countries suggests yes. Profits have been hard, if not impossible, to come by; many firms were struggling or collapsing; and layoffs and redundancies were the norm~\citep{ACCC2019}.
As \citet{Fenton2011} wrote in a paper centred on the UK, `News media are in crisis. The crisis is being managed by closing papers or shedding staff [and] these cuts are having a devastating effect on the quality of the news.' That was nearly a decade ago. Subsequent research suggests the situation has worsened significantly. In Australia, the commonly cited figure based on research by the journalists' union is that 3,000 journalism positions have been lost since 2011~\citep{Ricketson2020}. For instance, it is estimated that in 2011 news publisher Fairfax Media employed about 1,000 editorial staff across the \textit{Sydney Morning Herald}, \textit{The Age}, \textit{The Australian Financial Review}, and its Sunday papers, \textit{The Sun Herald} and \textit{The Sunday Age}. By mid-2017, however, half of those jobs were gone~\citep{Zion2018} (including the job of one of this paper's authors). And then the coronavirus wielded its scythe. Already, the reported impact of COVID-19 on journalism jobs has been devastating, with widespread closures and job losses, particularly in regional areas~\citep{Crerar2020}.

This research assesses the extent of the claimed `journalism crisis' in Australia by analysing labour market data from January 2012 to March 2020. 
To do this, we performed a quantitative analysis of two longitudinal data sets:
job advertisements (ads) for journalism jobs and the official Australian employment statistics.
This allowed us to measure longitudinally the demand for and supply of journalism jobs in Australia. 
Further, the breadth and detail of these data provided us with the opportunity to comprehensively assess the quality and characteristics of these journalism jobs.
Not only did we examine how key features of journalism jobs have changed -- such as salaries, location or years of experience -- but also how journalism skills required in Australia have evolved.
Additionally, the available data enabled us to measure the early effects of COVID-19 on journalism jobs.

Our findings confirm that there is a crisis in Australian journalism; a crisis that appears to have worsened during the early stages of the COVID-19 pandemic. 
However, the data also yields more granular findings, including three surprise findings. 
The first finding is that advertised journalism jobs only started to decline from 2016, not before. 
The second finding is that as the journalism jobs market became more volatile, gender inequity worsened: women journalists who remained were younger and worse paid than the men. 
And the third finding is that, according to our skill similarity calculations, generalist skills such as `Communications', `Public Relations', and `Social Media' became more important to journalism, as opposed to traditionally specialist journalism skills such as `Reporting', `Editing', and `Investigative Journalism'. 
These findings, together with others, reveal that the crisis in journalism is not only real, but in some ways more complex than was previously understood.

By implementing a data-driven methodology, we provide a comprehensive and longitudinal assessment of journalism jobs in Australia from January 2012 to March 2020. We tease out granular and specific trends, including the early impacts of the COVID-19 pandemic, the contrasting effects on regional and urban journalism jobs, and the gendered nature of ongoing impacts. And we analyse the underlying skills data to identify the skills sought in journalism jobs, and where people with journalism skills are likely finding alternate career paths.

\section{Relevant Literature \& Background}
\label{sec:related-work}
\textbf{Journalism jobs in crisis.}
If there is a crisis, the simple explanation is the Internet. 
(Putting aside COVID-19, to which we will return.) 
While digital channels have given journalism bigger audiences, they have also strangled income. 
Once, advertising funded journalism, but now advertising has largely migrated online. 
As the Australian Competition and Consumer Commission (ACCC) found in 2019, in the Final Report of its Digital Platforms Inquiry, `The reduction in advertising revenue over the past 20 years, for reasons including the rise of online advertising, appears to have reduced the ability of some media businesses to fund Australian news and journalism'. 
The ACCC cited Census data showing that `from 2006 to 2016, the number of Australians in journalism-related occupations fell by 9 per cent overall, and by 26 per cent for traditional print journalists (including those journalists working for print/online news media businesses)'. 
Further, the ACCC cited data provided by leading media companies showing that the number of journalists in traditional print media businesses fell by 20 per cent from 2014 to 2018 – a time of growth for Australia's population and economy~\citep{ACCC2019}.

However, the pressures on news media were not spread evenly. 
For instance, local news in particular bore the brunt. 
Between 2008 and 2018, 106 local and regional newspaper titles closed across Australia, representing a 15 per cent decrease in the number of such publications. 
As a result, 21 local government areas previously served by a newspaper were now without coverage, including 16 local government areas in regional Australia~\citep{ACCC2019}. These figures are mirrored in the US. 
In 2018, \citet{Abernathy2018} from the Hussman School of Journalism and Media at UNC released a report, `\textit{The Expanding News Desert}', which found that the US had lost almost 1800 papers since 2004, with 7112 remaining (1283 dailies and 5829 weeklies). 
This meant that the US lost roughly 20 per cent of its newspapers between 2004 and 2018. 
These closures included large dailies such as the \textit{Tampa Tribune} and the \textit{Rocky Mountain News}, but also many newspapers that had circulations of fewer than 5000 and served small, impoverished communities.

As the above research reveals, news media companies were under pressure, and journalism jobs were being cut.
There was some hope in the shape of new players entering the market and hiring journalists, including digital natives such as Vice and Buzzfeed.
However, in 2019 these two companies were among the many that announced significant staff layoffs~\citep{Goggin2019}. 
Worse, in 2020 Vice cut a further 155 jobs and Buzzfeed furloughed many of its workers without pay~\citep{Izadi2020}.
Furthermore, as Australia's ACCC notes, these publications `tend to employ relatively few journalists'~\citep{ACCC2019}. 
Even accounting for new arrivals, the number of journalism jobs in Australia continued to fall
(our own analysis in \textit{\nameref{sec:jobs-data-analysis}} also shows this trend), and as a result there were areas (including local government, local court, health and science issues) that journalism no longer covered adequately~\citep{ACCC2019}.

Further research has also revealed a clearer profile of the typical journalist, and also the typical journalist who loses his/her job. 
Drawing on 2017 data, one study found that journalism jobs internationally were largely filled by a young, inexperienced and itinerant workforce~\citep{Josephi2018}. 
Meanwhile, research suggests that it was journalists with extensive experience who were losing their jobs (at least in Australia)~\citep{Sherwood2018}. 
And those who lost their jobs faced decidedly uncertain futures. 
In longitudinal research tracking the post-journalism careers of Australian journalists who had been made redundant, many of those surveyed revealed they were experiencing job precarity~\citep{Zion2018}.
Further, a significant minority had moved into strategic communications or public relations~\citep{Zion2018}.
However, the flow of journalists into PR (and sometimes back again) is not new~\citep{carey1965communications,fisher2014watchdog, macnamara2014journalism, macnamara2016continuing} and our analysis also supports these previous results.

The nature of `journalism work' has also changed. 
Increasingly, scholars have sought to theorise journalism in terms of boundaries and blurring~\citep{carlson2015boundaries, loosen2015notion, ORegan2019, maares2020exploring}. 
The idea of blurred boundaries is intended to capture the ways in which journalism is increasingly difficult to define, and how traditional notions of journalism have been upended in the digital age~\citep{loosen2015notion}. 
Empirical work suggests that journalism is a fluid concept that now means many things, and that the definition of journalism is changing over time~\citep{bogenhold2013blurred}. 
For example, many contributors to social networks, including Instagrammers, can be considered to be creating work that is journalism~\citep{maares2020exploring}. 
As such, there is now no such thing as a typical journalist; rather, journalism is marked by diversity and heterogeneity rather than any unifying concept~\citep{bogenhold2013blurred}. 
The notion of blurred boundaries aligns with our findings regarding the way journalism jobs, and journalism skills, have been shifting. 
Indeed, \citet{carlson2016establishing} argues that journalism is uncertain, which means that scholars and audiences need to work towards clarifying both the value of journalism, and its meaning. 
Our research has been data-driven, analysing journalism jobs data according to the Australian occupational classifications of journalists (see Supplemental Material~\cite{appendix}) and based on their underlying skills in job ads data. 
Nonetheless, we suggest that our findings, coupled with previous research, have the potential to further inform how precisely journalism jobs in Australia have changed during this tumultuous period for the media industry.

\textbf{The impacts of COVID-19.}
There is a growing body of research into the impacts of COVID-19 on news and its audiences. 
Unsurprisingly, the research reveals that the outbreak of the global pandemic was accompanied by a marked upswing in news consumption in Australia~\citep{park2020covid}, the US~\citep{casero2020impact} and the UK~\citep{kalogeropoulos2020initial}. 
In Australia in 2019, 56 per cent of Australians accessed news more than once a day; by April 2020, three months after the first local case of COVID-19 was confirmed, that figure had jumped to 70 per cent~\citep{park2020covid}. 
Among other things, this increase involved audiences returning to television and legacy media in greater numbers~\citep{park2020covid, kalogeropoulos2020initial}. 
Soon, however, many people started avoiding news – and especially news about coronavirus - because it made them anxious~\citep{park2020covid, kalogeropoulos2020initial}. 
As \citet{kalogeropoulos2020initial} wrote following a survey of UK audiences conducted in May, `After an initial surge in news use, there has been a significant increase in news avoidance.'

Ultimately, COVID-19 gave rise to a paradox. 
The above surveys show that, as audiences sought out information to stay safe, there was a dramatic surge in the consumption of news - at least initially. 
At the same time, however, news outlets found it even harder to make money, as advertising dried up even further ~\citep{radcliffe2020covid, Doctor_Nieman, olsen2020communal}. 
With concerts cancelled and restaurants shuttered, promoters and restaurateurs had nothing to advertise, and the impacts on local and regional news were especially harsh~\citep{Doctor_Nieman}. 
On March 25, 2020, \textit{The Atlantic} ran a story under the headline, `The coronavirus is killing local news'~\citep{Steven_Waldman2020}.
The author urged people to subscribe: `Among the important steps you should take during this crisis: Wash your hands. Don't touch your face. And buy a subscription to your local newspaper.'
As one US media expert noted in late March, `Advertising, which has been doing a slow disappearing act since 2008, has been cut in half in the space of two weeks'~\citep{Doctor_Nieman}.

Even before COVID-19, the advertising crisis for journalism has been described not as a single black swan, but as a flock of black swans~\citep{Doctor_Nieman}. 
By one estimate, from 2006 to 2020, US newspapers lost more than 70 percent of their advertising dollars~\citep{Doctor_Nieman}. 
COVID-19 further cruelled advertising, compounding the strain on news media and the journalists they employ~\citep{olsen2020communal}.

In Australia, there were widespread closures and job losses before the pandemic, but COVID-19 compounded the problem. 
In late March, Rupert Murdoch's publishing business News Corp warned of `inevitable' job cuts and the closure of regional titles~\citep{Meade2020}. 
Soon afterwards, News Corp – Australia's biggest publisher - suspended the print editions of 60 Australian newspapers, including the \textit{Manly Daily} and \textit{Wentworth Courier} in Sydney, the \textit{Brisbane News} and the \textit{Mornington Peninsula Leader} in Victoria~\citep{Meade2020}.
In May, News Corp confirmed that more than 100 of its local and regional mastheads would either switch to digital only or disappear completely~\citep{Meade2020-100-cut}. 
These cuts came in the wake of a dramatic drop in advertising from the entertainment, restaurant and real estate industries, the titles' main revenue sources.
The global pandemic is ongoing, and its lasting impact on journalism remains to be seen. 
Our findings, drawn from data that runs until March 2020, are early and indicative rather than definitive.

\textbf{Job ads as a proxy for labour demand.} 
Job ads provide `leading' indicators of shifting labour demands as they occur, as opposed to the `lagging' indicators from labour market surveys. 
Consequently, job ads are increasingly used as a data source for analysing labour market dynamics~\citep{Markow2017-hu,Blake2019-ha}.
For instance, job ads data have also been used to assess labour shortages.
\citet{Dawson2019} defined a range of indicators to evaluate the presence and extent of shortages, such as posting frequency, salary levels, educational requirements, and experience demands. 
They also built a metric based on the forecasting error from Machine Learning models trained to predict posting frequency.
Intuitively, occupations experiencing high posting volatility are difficult to predict. 
Subsequent work showed these indicators to be predictive of labour shortages in the Australian Labour Market~\citep{Dawson2020a}.
In the present research, in \textit{\nameref{sec:jobs-data-analysis}}, we use a similar set of indicators to analyse labour demand for journalists.
Further details on job ads data are provided in the Supplemental Material~\cite{appendix}.

\textbf{Analysing journalism jobs.}
Journalism jobs have previously been analysed using job ads. 
\citet{young2018} collected and assessed how Australian media outlets defined journalism job positions when hiring journalists from November 2009 to November 2010. The authors used a content analysis methodology and manually labelled data fields, such as employer, educational qualifications, job responsibilities, experience requirements, location, work hours, media platform, skill demands, job title, and any other miscellaneous information. The authors found that journalism was not a high priority during this period; instead employers advertised four times as many job ads for sales, marketing, and advertising positions.

More recently, \citet{guo2019} conducted content analysis on 669 journalist job announcements from US media organisations from 1 July to 31 December 2017. The authors' objective was to define, compare, and analyse the journalists' expertise requirements as expressed through job ads. To achieve this objective, the authors manually reviewed and codified job vacancies. This research found that `multi-skilled' journalists are experiencing higher levels of demand. The authors also found that journalists' ability to flexibly adapt to changing situations was a characteristic of growing importance. 
These studies, while significant, are relatively limited in scope.
In this paper, we analyse a nine-year dataset of job ads which allows us to uncover longitudinal dynamics of journalism jobs.

Historic employment levels of journalists in Australia have also been analysed by \citet{ORegan2019}. The authors used five-yearly census data and found that not only has the advent of digital platforms coincided with the decline of many types of journalists (for example, `Print', `Radio', `Television' and `Editors'), but employment has shifted into related professions, such as `Authors' and `Public Relations'. O'Regan and Young's paper built on earlier research by \citet{higgs2007australia}. Our research complements the findings of \citet{ORegan2019}, providing additional labour demand detail from job ads data while also matching it with labour supply data from employment statistics.

\textbf{Limitations of job ads data.} 
Job ads data are an incomplete representation of labour demand. 
Some employers use traditional forms of advertising for vacancies, such as newspaper classifieds, their own hiring platforms, or recruitment agency procurement. 
Furthermore, anecdotal evidence reveals that some vacancies are filled informally, using channels such as word of mouth, professional networks and social media. 
Job ads data also over-represent occupations with higher-skill requirements and higher wages, colloquially referred to as `white collar' jobs~\citep{Carnevale2014-xc}. 
Finally, just because a job is advertised, does not mean that the position will be, or has been, filled.
Despite these shortcomings, job ads provide extremely rich information for what employers are demanding in near real-time; including information that cannot be gathered from employment statistics. Given the sample size of journalism job ads available and the detailed skills extracted in the data set, we are confident that the journalism job ads used for this research provide a useful indication of journalism labour demand.

\textbf{Employment statistics and occupational standards.} Employment statistics provide data on populations employed in standardised occupational classes. Occupations in Australia correspond to their respective occupational classes according to the Australian and New Zealand Standard Classification of Occupations (ANZSCO)~\citep{ANZSCO2013}. 

There are significant shortcomings to analysing occupations within ANZSCO categories.
Official occupational taxonomies (like ANZSCO) are often static and are rarely updated, therefore failing to capture emerging skills, which can misrepresent the true labour dynamics of particular jobs. 
For example, the occupational class of `Print Journalist' has been a constant in Australian occupational statistics. 
Yet, the underlying skills of a `Print Journalist' have changed dramatically in recent decades. 

To overcome the above-stated limitations, in our data construction, we leveraged the Burning Glass Technologies (BGT -- the job ads data source) occupational ontology together with the ANZSCO ontology. 
We also used the rich skill-level information from job ads that are missing from occupational employment statistics to build an encompassing journalism job ads dataset.
\section{Data \& Methods}
\label{sec:data-methods}
\subsection{Data Sources}
This research used both labour demand and labour supply data to analyse journalism jobs. 
On the labour demand side, we used a detailed dataset of over 8 million Australian job ads, spanning from January 2012 to March 2020. 
These data were generously provided by Burning Glass Technologies\footnote{BGT is a leading vendor of online job ads data. \texttt{https://www.burning-glass.com/}} (BGT).
For labour supply data, we leveraged official employment statistics~\citep{Australian_Bureau_of_Statistics2019-sv} and salary levels~\citep{ABS-earnings} provided by the Australian Bureau of Statistics (ABS) over the same period. 
These data sources provide longitudinal employment and salary information that have been disaggregated by gender, location, and types of employment (full-time and part-time). 
Further details of data sources and data construction are provided in the Supplemental Material~\cite{appendix}.
While there are nuances to `journalism work' and the requirements of journalism jobs have evolved over time~\citep{macnamara2016continuing, ORegan2019, maares2020exploring}, 
this research defines journalism jobs by the official ANZSCO standards~\citep{Abs2019-journalists-unit}.

\subsection{Skill Similarity}
\label{subsec:skill-similarity}
To analyse the underlying journalism skills within occupations, we implemented a skill similarity methodology adapted from \citet{Alabdulkareem2018-jl} and then by \citet{Dawson2019} to calculate the pairwise similarities between skills from job ads.

Two skills are similar when the two are related and complementary, i.e. the two skills in a skills-pair support each other. 
For example, `Journalism' and `Editing' have a high pairwise similarity score because together they enable higher productivity for a journalist; whereas `Journalism' and `Oncology' have a low similarity because
they are seldom required together.
We measured the similarity of skill-pairs based on their co-occurrence patterns in job ads, while accounting for skill ubiquity and specialisation.
To capture how journalism skills have changed over time, we measured skill similarity during calendar years.

Formally, given $J$ as the set of job ads posted during a specific calendar year, we measured the similarity between two skills $s$ and $s'$ as:
\begin{equation} \label{eq:theta}
 \theta(s, s') = \frac{\mathop{\sum}\limits_{j'\in J}e(j,s) e(j,s')}
 {max \left( \mathop{\sum}\limits_{j'\in J}e(j,s), \mathop{\sum}\limits_{j'\in J}e(j,s') \right)}
\end{equation}
where $j$ and $j'$ are individuals jobs ads from the set $J$, and $e(s, j) \in \{0,1\}$ measures the importance of skills $s$ for job $j$ using theory from Trade Economics~\citep{Hidalgo2007-qk}.
Skills $s$ and $s'$ are considered highly complementary if they commonly co-occur and are both `important' for the same job ads.
Finally, $\theta(s, s') \in [0, 1]$, a larger value indicates that $s$ and $s'$ are more similar, and it reaches the maximum value when $s$ and $s'$ always co-occur (i.e. they never appear separately).

We build the top yearly lists of journalism skills by computing $\theta(Journalism, s)$ -- i.e. the similarity between the skill `Journalism' and each unique skill that occurs for each year from 2014-2018. 
The yearly top 50 skills most similar to `Journalism' are shown in the Supplemental Material~\cite{appendix} together with the full details of the $\theta$ measure.

Finally, we determined the occupations with the highest levels of skill similarity to the top journalism skills uncovered from above. 
We propose $\eta$, the \emph{`Journalism Skill Intensity'}, for each standardised BGT occupation, defined as percentage of journalism skills relative to the total skill count for the job ads related to an occupation $o$. 
Formally:
\begin{equation} \label{eq:skill-intensity}
	\eta(o, \mathcal{D}) = \frac{\mathop{\sum}\limits_{j \in \mathcal{O}, s \in \mathcal{D}} x(j, s)}{\mathop{\sum}\limits_{j \in \mathcal{O}, s' \in S} x(j, s')}
\end{equation}
\noindent
where $\mathcal{D}$ is the set of journalism skills, and $\mathcal{O}$ is the set of job ads associated with the occupation $o$. This method allowed us to adaptively select occupations based on their journalism skill intensities.
\section{Jobs Data Analysis and Results}
\label{sec:jobs-data-analysis}
In this section, we conducted a data-driven analysis of journalism jobs in Australia based on job ads data and official occupational statistics.
First, we longitudinally examined key features of jobs data, such as employment levels, job ads posting frequency, salaries, and posting frequency growth and predictability level.
We also analysed how the underlying skills of journalists had changed over time, and which skills and occupations grew in similarity to journalism.

\subsection{Posting Frequency \& Employment levels}

\label{subsec:posting-freq-employment}

\begin{figure}[htbp]
 	\centering
	\newcommand\mywidth{0.49}
	\includegraphics[width = \mywidth\textwidth]{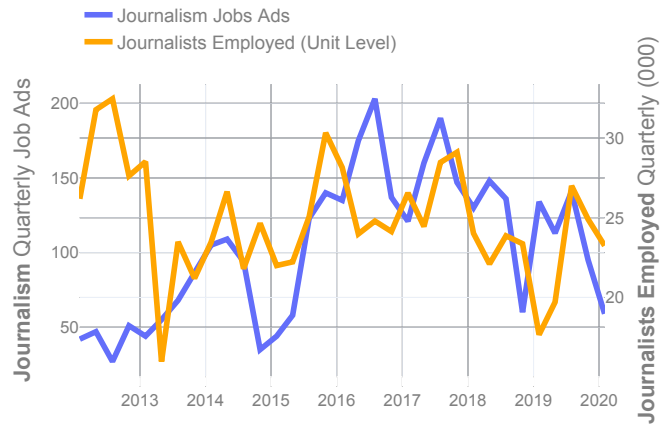}\caption{
	 Quarterly posting frequency of journalism job ads (see Sec.~\textit{\nameref{sec:data-methods}}) and employment levels of `Journalists \& Other Writers' at the ANZSCO Unit level (000's) from Jan 2012 to Mar 2020. 
}
	\label{fig:quarterly-posts-unit-emp}
\end{figure}

In Australian journalism, 2012 is considered a watershed year. 
An estimated 1,500 journalists were made redundant, the majority of those from Australia's two largest print companies, Fairfax Media (now Nine Entertainment) and News Limited (now News Corp Australia)~\citep{zion2016}.
The severity of this industrial shock can be observed in \cref{fig:quarterly-posts-unit-emp}. 
Against the left y-axis, the blue line shows quarterly job ads posting frequency for journalism jobs. 
As the graph depicts, posting frequency for journalism job ads experienced extremely low levels in 2012 until 2013, when they began to increase. 
The volume of vacancies increased until mid-2014, before plummeting in late-2014 to the levels last seen in 2012. 
From 2015, journalism job ads experienced strong growth, reaching a peak in mid-2016.
Since then, journalism job ads have trended downward until the first quarter of 2020 (end of available data for job ads), albeit with volatile peaks and troughs. 
In summary, the data shows that journalism job ads had not been in freefall since 2012. Rather, there was erratic growth in journalism job ads until a peak in 2016, followed by erratic decline. 

Similarly, employment levels underwent immense volatility from 2012 to 2013. 
Against the right y-axis of \cref{fig:quarterly-posts-unit-emp}, the orange line shows the number of quarterly employed for `Journalists \& Other Writers' at the ANZSCO Unit level. 
Employment levels peaked in mid 2012, before dramatically dropping in early 2013. 
This is an effect of the mass journalist redundancies made in 2012, given that employment statistics are `lagging indicators' and it takes time for labour markets to reflect changes in occupational statistics.
Early 2013 marked the lowest point of journalist employment seen in this time-series. 
As also observed in job ads data, journalist employment levels grew until 2016-2017 and has since trended downwards, exhibiting volatile quarterly changes through to the first quarter of 2020.

\begin{figure*}[htbp]
	\centering
	\newcommand\mywidth{0.98}
	\includegraphics[width = \mywidth\textwidth]{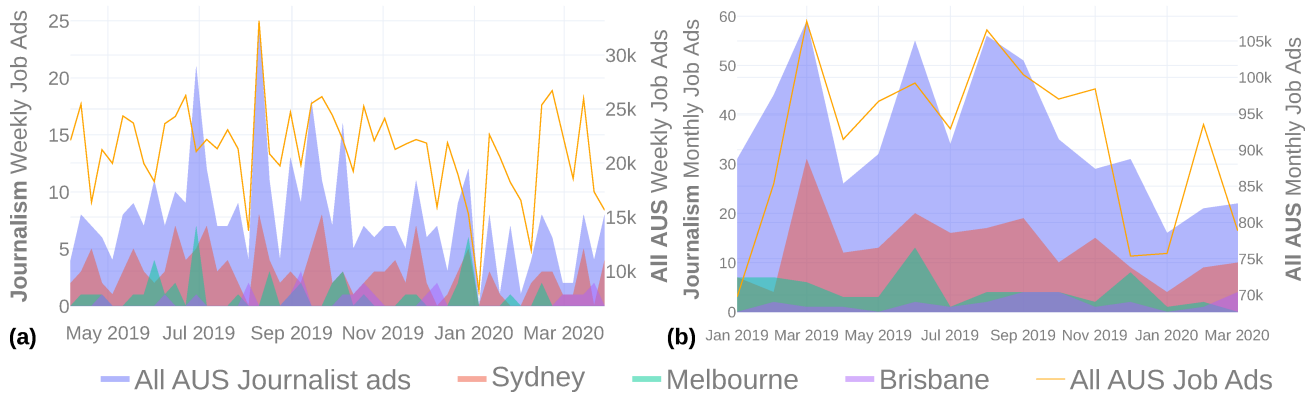}\caption{
		\textbf{Posting frequency for journalism jobs during the early stages of the COVID-19 crisis in Australia and its major cities}:
	 \textbf{(a)} Weekly posting frequency volumes for journalists and all Australian job ads between April 2019 and March 2020.
	 Both decreased as the early stages of the COVID-19 crisis hit;
	 \textbf{(b)} Monthly posting frequency for journalists were down 63 per cent when comparing March 2019 to March 2020. 
	 This was significantly higher than all Australian job vacancies, which was down 37 per cent over the same period.
	 }
	\label{fig:covid-posting-freq}
\end{figure*}

\textbf{COVID-19 and journalism jobs.} The early effects of COVID-19 were apparent in the posting frequency of job ads in Australia. This was the case for most occupations, including journalists. Higher vacancy rates typically mean higher levels of labour demand by employers, which is a critical component of healthy labour markets.
As \cref{fig:covid-posting-freq} highlights, vacancy volumes declined for both journalism jobs and at aggregate levels in Australia. 
Since mid-February 2020, weekly posting frequency had decreased across all Australia job ads, as seen in \cref{fig:covid-posting-freq}a. 
Such a decline this early in the year is atypical. 
As \citet{Dawson2020} show, the frequency of job ad postings follow a yearly seasonal pattern, with late February and early March typically being a period of upward trend growth. However, late February and early March 2020 coincided with the international outbreak of COVID-19. During this period, the Australian government instituted widespread quarantine and social distancing measures, which significantly constrained economic activity~\citep{Boseley2020}. 
The impacts of these COVID-19 containment laws are starkly apparent in \cref{fig:covid-posting-freq}b. 
Posting frequency for journalism jobs were down 63 per cent when comparing March 2019 volumes to March 2020. This was significantly higher than the aggregate market of all Australian job ads, which was down 37 per cent over the same period. 
\cref{fig:covid-posting-freq}b shows that Melbourne appeared to be the city hardest hit, recording no journalism job ads in March 2020 and only 3 posts for the first quarter of 2020, even before the major lock-downs instituted for Melbourne in August 2020. Clearly the pandemic had an early and damaging effect on the journalism jobs market.

\begin{figure}[htbp]
 	\centering
	\newcommand\mywidth{0.49}
	\includegraphics[width = \mywidth\textwidth]{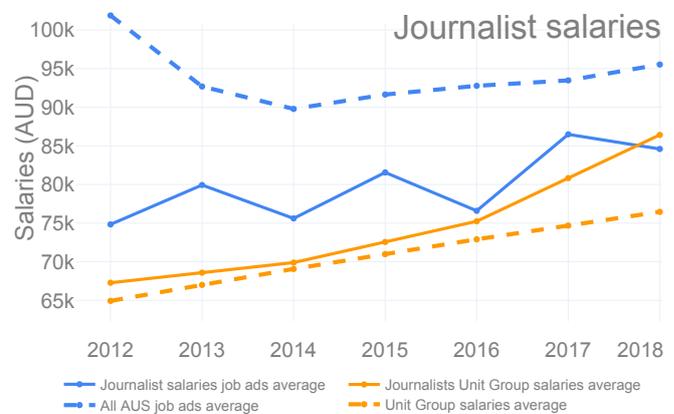}
	\caption{
	 Journalist salaries (\textcolor{blue}{solid blue line}) increased according to job ads data, but remained below market average levels (\textcolor{blue}{dashed blue line}). 
	 However, according to ABS data, `Journalists \& Other Writers' (ANZSCO Unit level, \textcolor{orange}{solid orange line}) earned a growing wage premium above the market average (\textcolor{orange}{dashed orange line}).
	}
	\label{fig:salaries}
\end{figure}

\subsection{Salaries}
We compared salaries extracted from job ads with ABS reported wage data for `Journalists and Other Writers'\footnote{ABS wage data is reported biennially, with the latest reporting year being 2018. Therefore, wage values in the `odd' years in between the reporting periods were interpolated, calculated as the mean of the previous and the subsequent years.}.
\cref{fig:salaries} reveals two main findings regarding journalist salaries. 
First, according to job ads data, journalists attracted considerably lower annual wage levels (solid blue line) than the market average (dashed blue line).
As of 2018, job ads indicated that journalists earned approximately AU\$10,000 less than the market average.
These findings, however, are somewhat contrary to the wage earnings data collected by the ABS~\citep{ABS-earnings}, according to which `Journalists and Other Writers' (solid orange line) had been earning a growing wage premium over the market average (dashed orange line) since 2014.
This discrepancy can be explained by the fact that job ads data tend to over-represent occupations in the `Professional' and `Manager' classes~\citep{Carnevale2014-xc}, which typically attract higher wages.
As a result, the average salary levels from job ads data (dashed blue line) were about AU\$20,000 higher than average salary levels from ABS data (dashed orange line), from 2014 to 2018.
However, the salary levels for journalists were very similar when comparing across the two data sources.

\cref{fig:salaries} yields a second observation:
journalist salary levels increased in both absolute and relative terms compared to average market levels, between 2012 to 2018 in both data sources. 
More importantly, the relative salary growth of journalists exceeded the market averages, during the period studied.

\begin{figure}[tbp]
 \centering
	\newcommand\mywidth{0.48} 
		\includegraphics[width = \mywidth\textwidth]{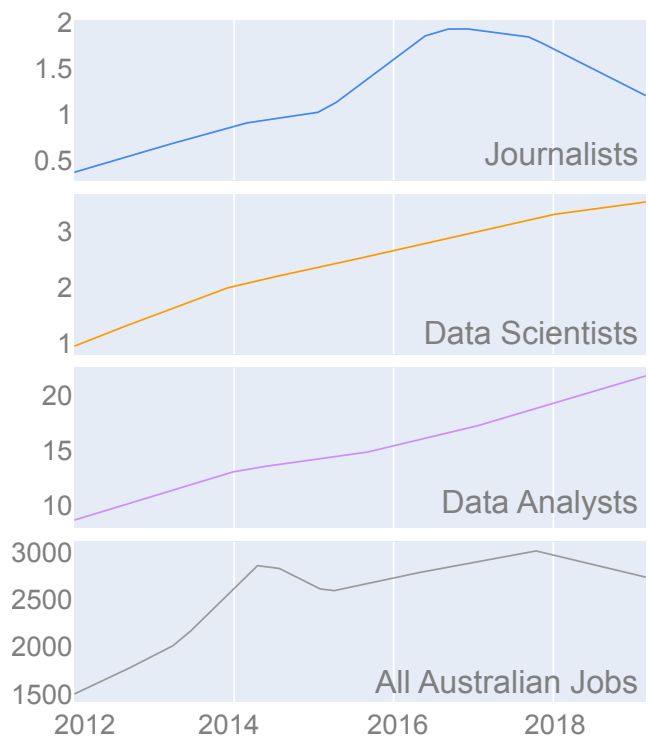}
	\caption{
		\textbf{Trend lines of posting frequency} for `Journalists', `Data Scientists', `Data Analysts', and `All Australian job ads'. 
		Posting frequency for `Journalists' trended downwards since 2016.
	 }
	\label{fig:trendlines}
\end{figure}

\subsection{Trend Analysis \& Predictability}
\label{subsec:trend-predictability}

\textbf{Posting trends.}
We constructed an auto-regressive Machine Learning model to predict posting frequency of journalism job ads in Australia~\citep{taylor2018forecasting}. 
The model accounts for long term trends, seasonality patterns and external events (see the Supplemental Material~\cite{appendix} for technical details).
We isolated the posting frequency trend component and, in \cref{fig:trendlines}, plotted it comparatively for `Journalists' against two occupations that have experienced high levels of labour demand, `Data Scientists' and `Data Analysts', as well as against the aggregated market trend.
Visibly, journalism jobs experienced varying degrees of growth until mid 2016, at which point growth plateaued, and started to decline. 
From the end of 2017 until 2019, the trend for journalism job ads has heavily decreased, even when compared to the aggregate market, which also shows a more modest decrease during the same period.
`Data Scientists' and `Data Analysts' consistently grew throughout the entire period.

\begin{figure}[htbp]
 	\centering
	\newcommand\mywidth{0.48}
	\includegraphics[width = \mywidth\textwidth]{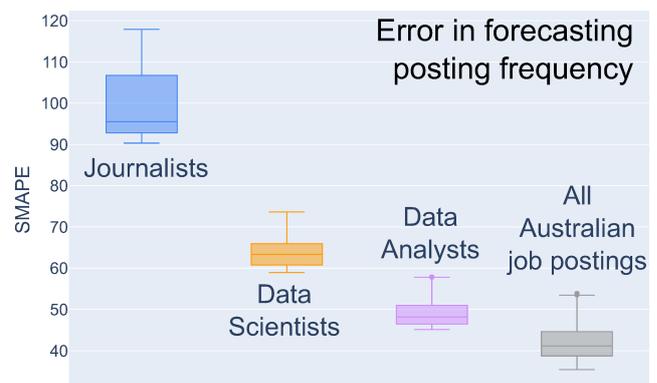}
	\caption{
	 \textbf{(a)} Predictability comparison of temporal posting frequency highlighting the difficulties of predicting journalism job ads and their volatility.
	}
	\label{fig:predictability}
\end{figure}

\begin{figure*}[tbp]
	\centering
	\newcommand\myheight{0.17}
	\subfloat[]{
		\includegraphics[height = \myheight\textheight]{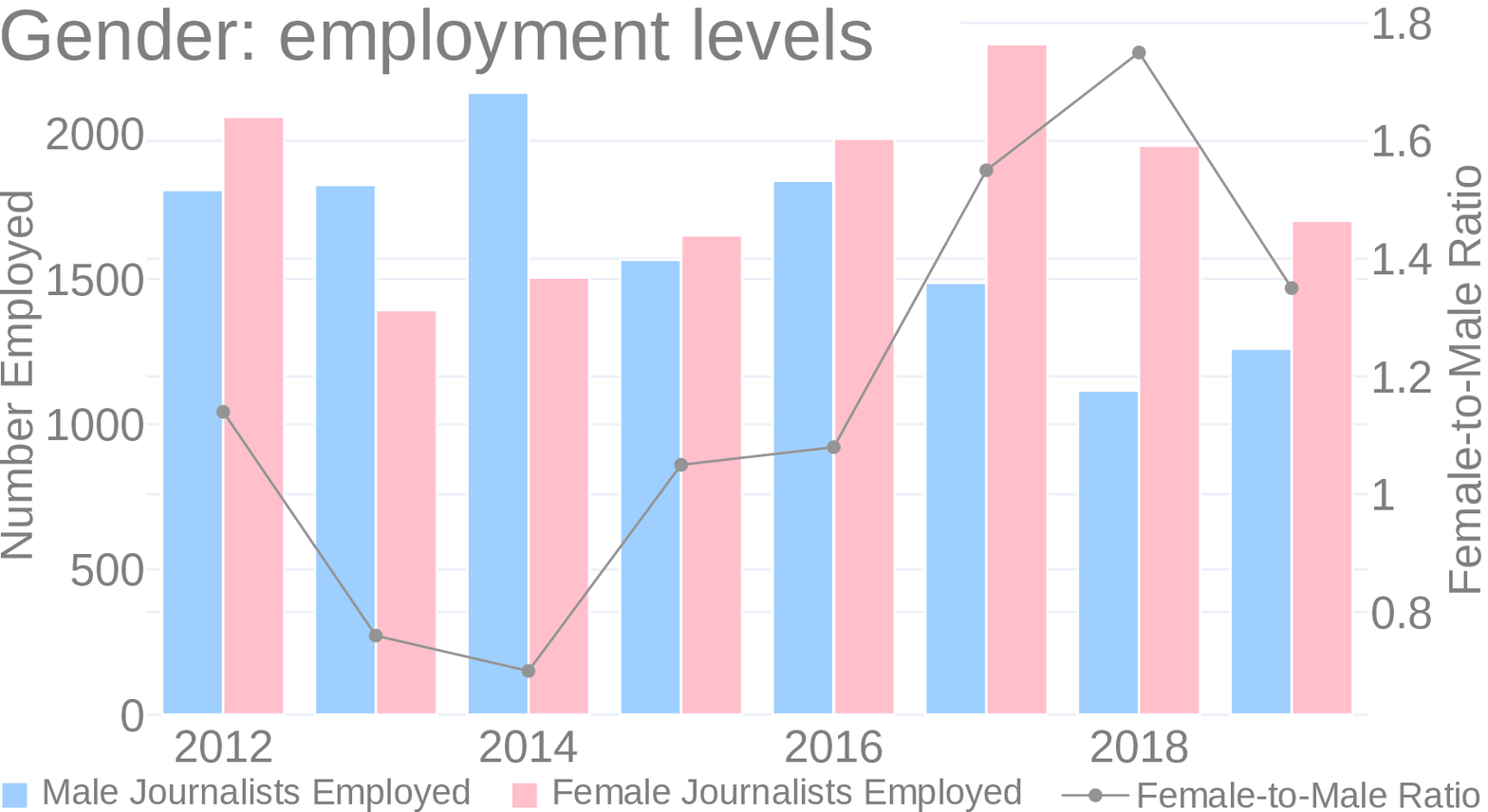}\label{subfig:gender-employed}
	}
	\subfloat[]{
		\includegraphics[height = \myheight\textheight]{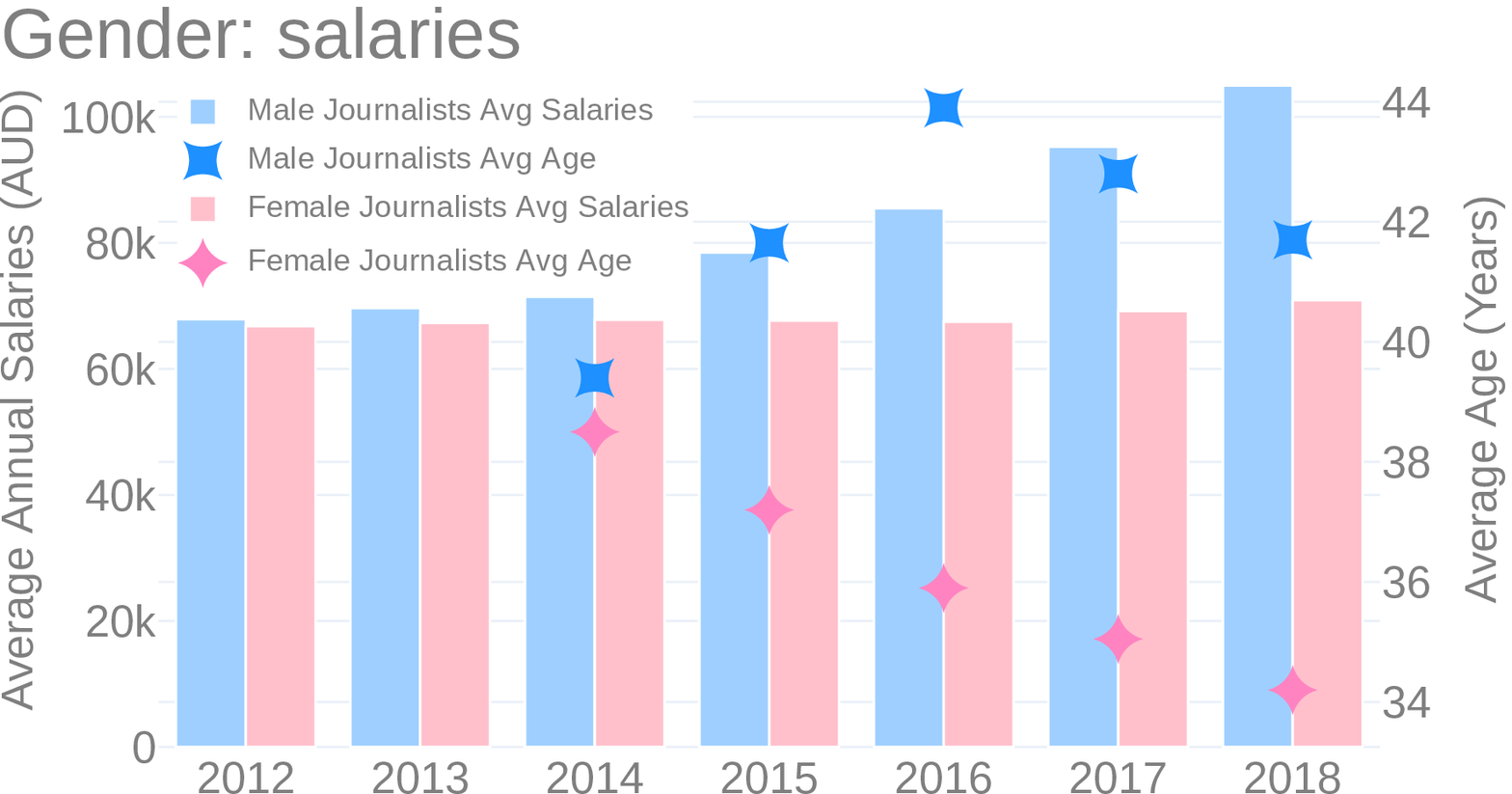}
		\label{subfig:gender-salaries}
	}
	\caption{
		\textbf{Journalist employment levels and salaries by Gender}:
	 \textbf{(a)} Since 2015, the employment ratio of female-to-male journalists increased;
	 \textbf{(b)} Wage inequality increased between males and females in the `Journalists \& Other Writers' Unit group. 
	 This was at the same time that the average age of journalists decreased for females and increasing for males since 2014.
	 }
	\label{fig:gender}
\end{figure*}

\textbf{Quantify labour demand volatility.}
When constructing Machine Learning models, it is standard procedure to use error metrics to evaluate the prediction accuracy. 
Volatility in posting volumes inherently lead to lowered prediction performance and higher error values.
Here we use the prediction error measured using the `Symmetric Mean Absolute Percentage Error'~\citep{Scott_Armstrong1985-tt, makridakis1993accuracy} as a proxy for the volatility of labour demand for different occupations (see the technical section in the Supplemental Material~\cite{appendix} for more details).

\cref{fig:predictability} shows the prediction performance for three occupations (`Journalists', `Data Scientists', `Data Analysts') and for the volume of `All Australian job postings'.
We used a sliding window approach to obtain multiple predictions (see the Supplemental Material~\cite{appendix}) that we aggregated as boxplots.
The higher the error score on the vertical axis, the lower the predictive abilities for that occupation. 
As \cref{fig:predictability} reveals, predicting the daily posting frequency of journalism jobs was consistently more difficult than for the other occupations, and the market as whole.
`Data Scientists', an occupation undergoing strong relative growth, is also showing a high prediction error compared to the market as a whole, indicative of experiencing a degree of volatility.
However, it was not nearly commensurate to the predictive difficulties, and volatility, of journalists. 
This was true from 2012 to 2019, and has become worse in 2020 with the spread of COVID-19.

\begin{figure*}[tbp]
	\centering
\newcommand\myheight{0.23}
	\subfloat[]{
		\includegraphics[height = \myheight\textheight]{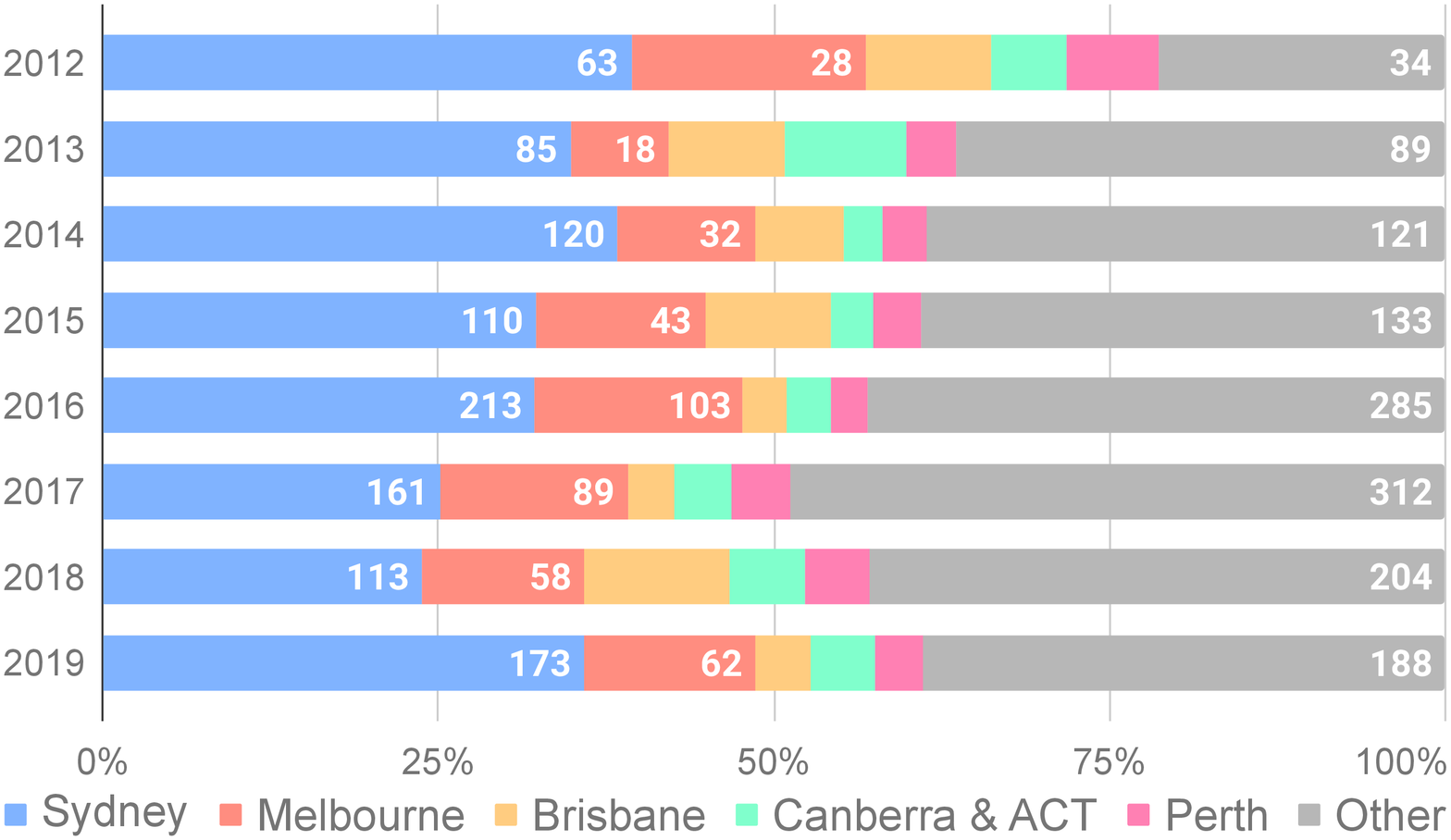}\label{subfig:cities}
	}
	\subfloat[]{
		\includegraphics[height = \myheight\textheight]{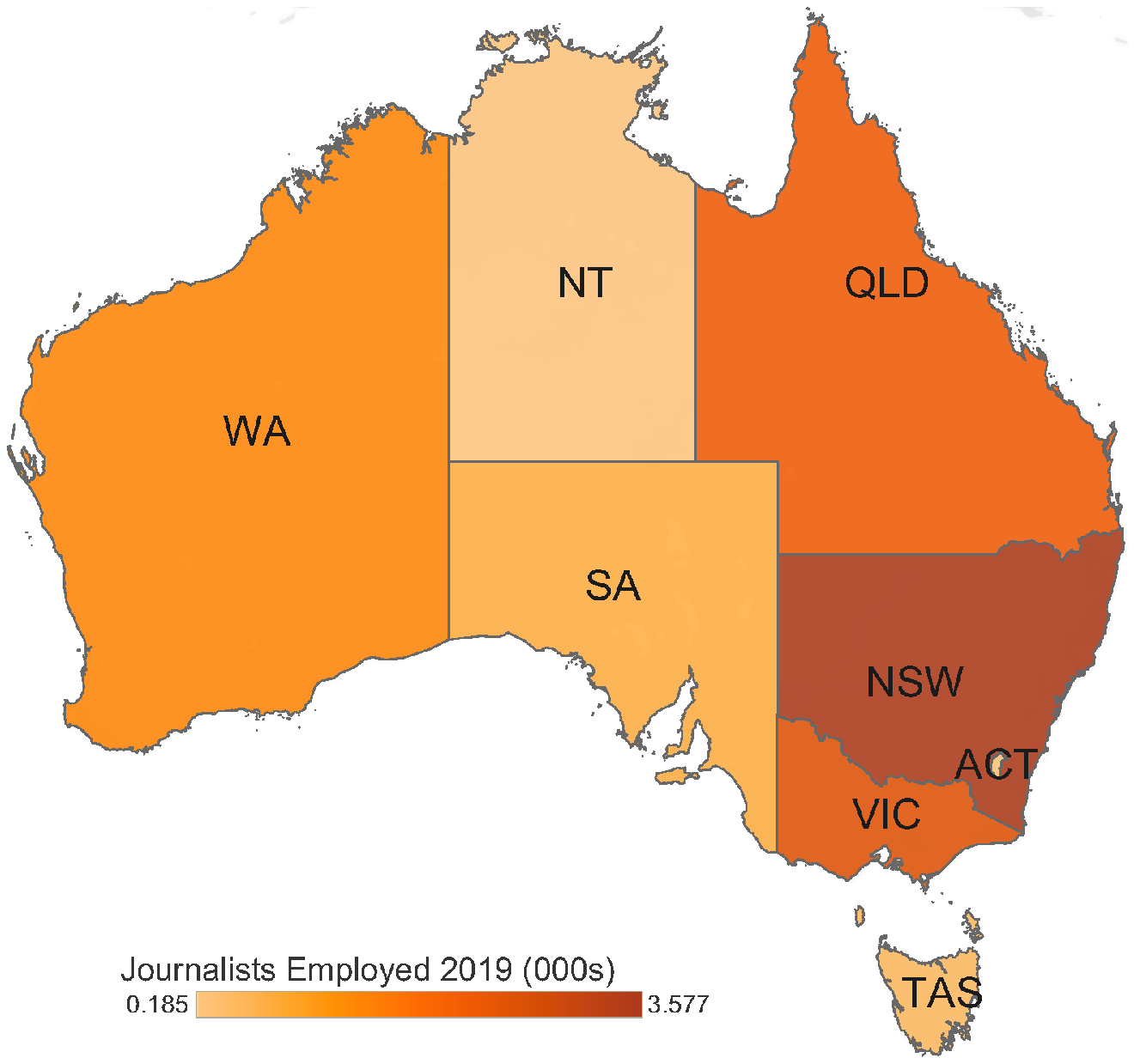}
		\label{subfig:employed-states}
	}
	\caption{
		\textbf{Location of journalists in Australia}:
	 \textbf{(a)} Posting frequency for journalism jobs decreased in major Australian cities, in relative terms; 
	 \textbf{(b)} As of 2019, the majority of journalists in Australia were employed in New South Wales, Victoria, and Queensland, respectively.
	}
	\label{fig:location}
\end{figure*}

\begin{figure*}[tbp]
	\centering
	\newcommand\mywidth{0.49} 
	\subfloat[]{
		\includegraphics[width = \mywidth\textwidth]{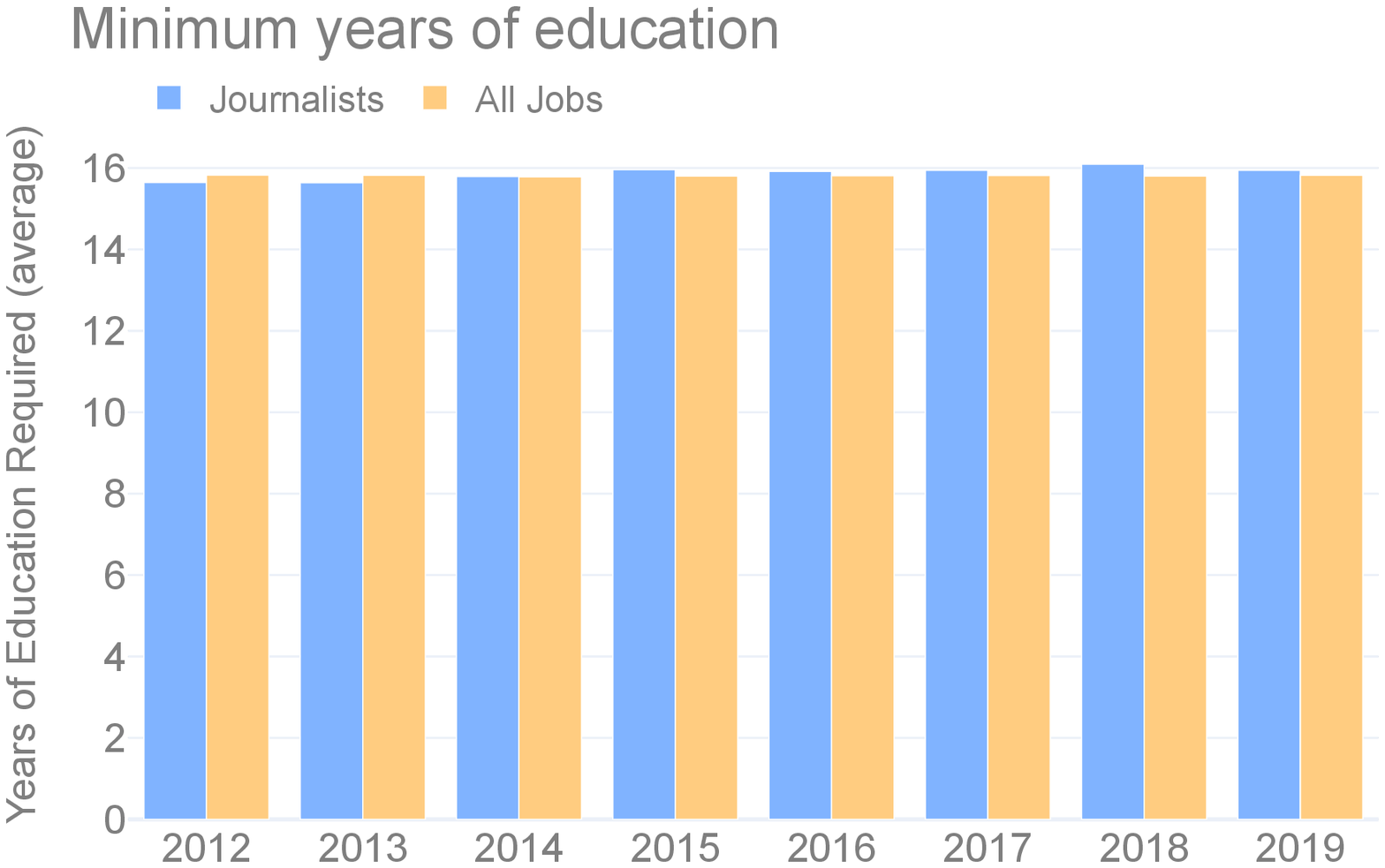}\label{subfig:education}
	}\subfloat[]{
		\includegraphics[width = \mywidth\textwidth]{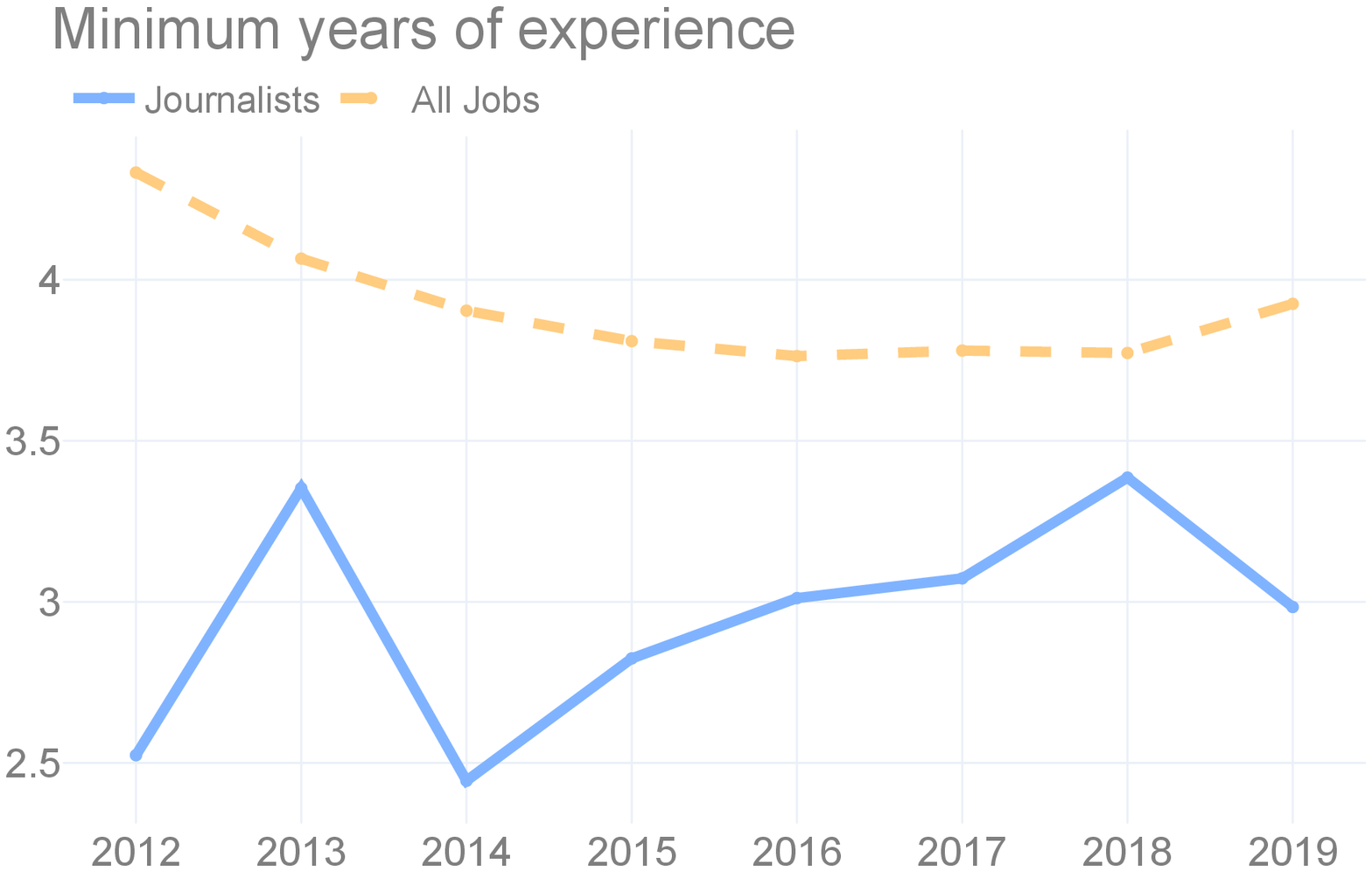}
		\label{subfig:experience}
	}
	\caption{
	 \textbf{(a)} Years of Education demanded by employers from job ads were consistent with the market average; 
	 \textbf{(b)} Years of Experience required by employers consistently remained below the market average, according to job ads. 
	 However, this gap had closed since 2014.
	 }
	\label{fig:education-experience}
\end{figure*}

\subsection{Gender}
There have been growing gender differences of employed journalists in Australia~\citep{north2016gender, north2009gendered} and across the world~\citep{hanitzsch2019worlds}; the data presented in this research reinforces these previous findings. 
\cref{subfig:gender-employed} shows that the ratio of female employed journalists increased relative to male journalists (ANZSCO Unit Level)~\citep{Australian_Bureau_of_Statistics2019-sv}. 
In 2014, the female-to-male employment ratio was 0.7. 
In 2018, the proportion more than doubled, with almost 1.8 female journalists employed for every male journalist. 
It then declined in 2019 to 1.35, but this proportion was still almost double that of 2014.

\cref{subfig:gender-salaries} also shows that wage inequality between female and male journalists worsened from 2014 to 2018~\citep{ABS-earnings}. 
Since 2014, the annual salaries for female journalists increased by only AU\$3,000, whereas annual salaries for male journalists increased by more than AU\$30,000.
Male journalists thus experienced an average wage growth that was ten times greater than female journalists from 2014 to 2018.

There were also changing age demographics of employed journalists during the studied period. 
The markers on \cref{subfig:gender-salaries} highlight the average age of journalists by gender, per year.
Male journalists were getting older, their average age increasing by two years from 2014 to 2018. 
Female journalists, however, were steadily getting younger. 
The average age for female journalists decreased by more than four years from 2014 to 2018.

\subsection{Location}
\cref{fig:location} plots the location and volume of employed journalists in Australia.
\cref{subfig:cities} shows the absolute and relative number of job ads posted for each of the capital cities, and outside them, and \cref{subfig:employed-states} shows the location of employed journalists per state.
Unsurprisingly, Sydney and Melbourne, the respective capital cities of New South Wales (NSW) and Victoria (VIC), consistently had the highest job ad posting frequencies. 
However, the relative share of job ad posting frequency in Australian capital cities had shrunk in later years, with \cref{subfig:cities} showing an increase outside of major cities, both in relative and absolute terms. 
This trend reached a peak in 2017, when less than 50 per cent of all journalist job ads were for positions inside capital cities.
A small rebound followed, and in 2019 Sydney commanded approximately one-third of all journalism job ads.

\subsection{Education \& Experience}
\cref{subfig:education,subfig:experience} show respectively the number of years of formal education required for journalists, and the experience requirements (both per year, extracted from job ads data).
The education requirements consistently remained at market average levels, with journalists required to possess a Bachelor-level degree (approximately 16 years of education).

By contrast, the experience requirements were more variable. 
Since 2012, employers required fewer years of experience from journalists than was required in the market generally. However, the gap narrowed. In 2019, employers demanded of journalists, on average, half of an additional year of experience compared to 2014. 
This countered the general market, where employers' demands trended downward from 2012 to 2019. 

\subsection{Employment Type}
\begin{figure}[htbp]
 \centering
	\newcommand\mywidth{0.49} 
	\includegraphics[width = \mywidth\textwidth]{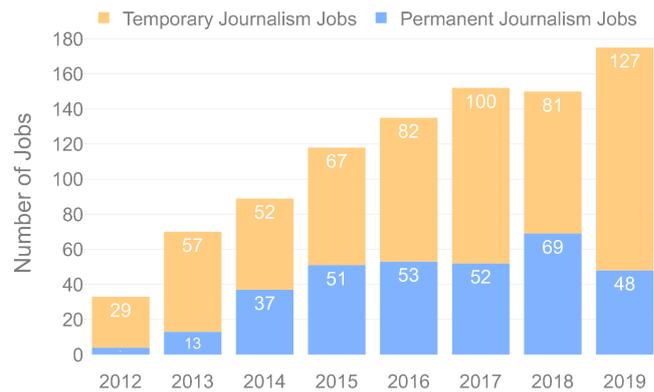}
	\caption{
	 Temporary positions represented the majority of journalism job ads in Australia. 
}
	\label{fig:temp-vs-perm}
\end{figure}

Casual and temporary work have become more commonplace in Australia~\citep{Gilfillan2018}, and we study if this is also the case for Australian journalism jobs.
In \cref{fig:temp-vs-perm} we plot the number of permanent and temporary journalism jobs, per calendar year -- in the job ads data, jobs are classified as either `Permanent', `Temporary', or not specified according to the content of the job descriptions.
The number of `Temporary' journalism jobs had increased in absolute terms since 2012, which made up the majority of all journalism ads in every year.
It is noteworthy too that the share of `Permanent' journalism vacancies had also increased since 2012. 
However, this trend should be interpreted with a degree of scepticism as only $\sim$ 50 per cent of all journalism job ads specified whether the roles advertised were permanent or temporary.

\subsection{Journalism Skills}

\textbf{Growing demand for journalism skills.} 

\begin{figure}[htbp]
 \centering
	\newcommand\mywidth{0.47} 
	\includegraphics[width = \mywidth\textwidth]{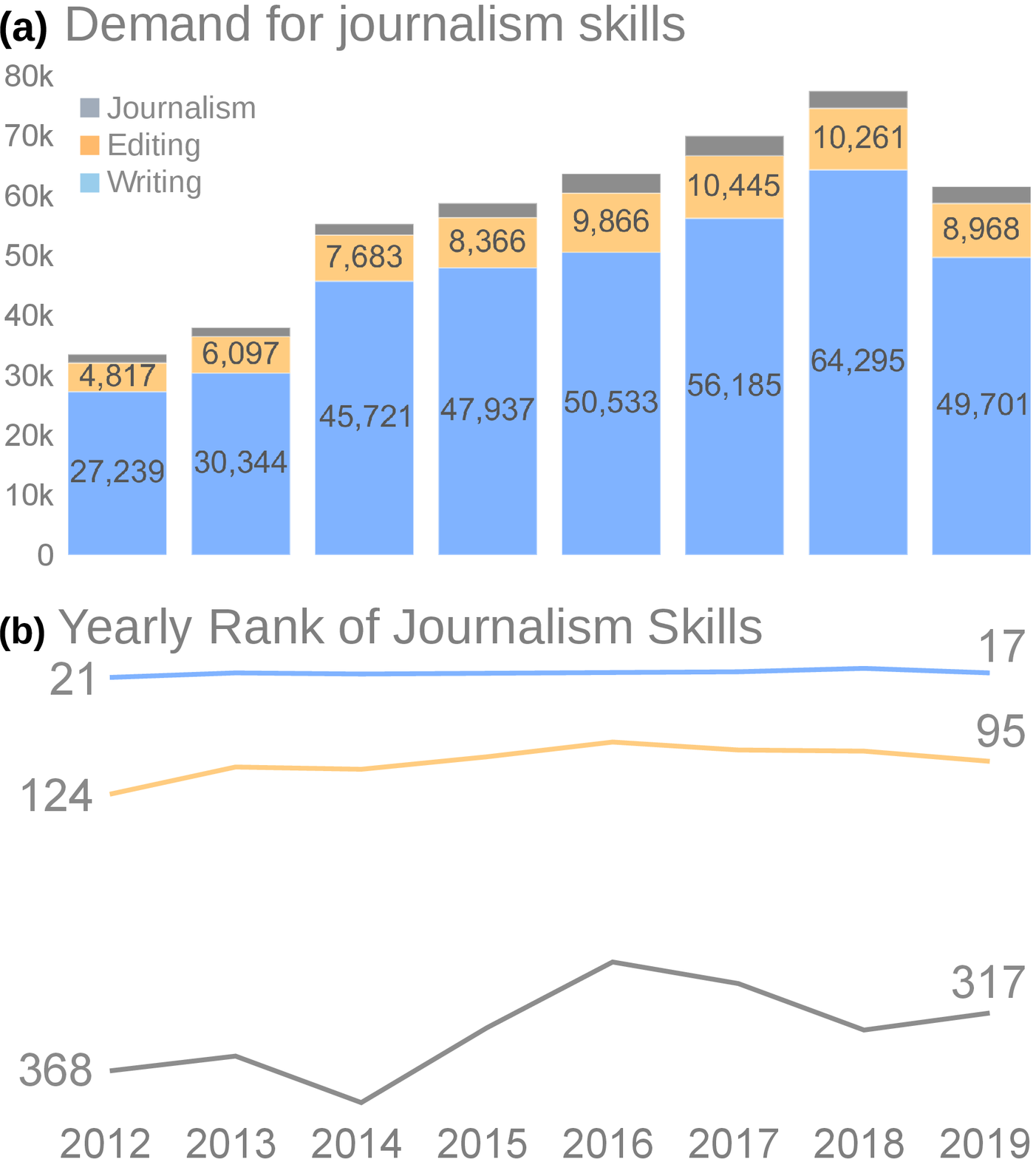}
	\caption{ 
		The absolute posting frequency \textbf{(a)} and relative yearly rank \textbf{(b)} of three major journalism skills increased between 2012 and 2019.
	}
	\label{fig:rank-count}
\end{figure}

Here, we analysed how the demand for some fundamental journalism skills changed over time. 
First, we selected three traditionally important skills to journalists that appear in job ads:
 (1) `Journalism', (2) `Editing', and (3) `Writing'. 
These skills were then counted across all job ads in Australia, regardless of their occupational class. 
While \cref{fig:trendlines} shows that labour demand for journalists has decreased since 2016, \cref{fig:rank-count}a presents the more nuanced story, showing that the posting frequency for each of these core journalism skills increased from 2012 to 2019, with 2018 to 2019 being the first yearly decline. 

The relative rankings of these three skills also increased. 
For each year, we counted the posting frequency of each unique skill that appears in job ads. 
We then ranked these skills by posting frequency as a proxy for labour demand. 
\cref{fig:rank-count}b shows that the rankings of all three of these fundamental journalism skills had improved from 2012 to 2019. 
In other words, not only did the posting frequency of these three journalism skills increase in job ads over these eight years, but their importance relative to all other skills also increased.

\label{subsec:journalism-skills}

\begin{figure*}[htbp]
	\centering
	\newcommand\myheight{0.27} 
	\subfloat[]{
		\includegraphics[height = \myheight\textheight]{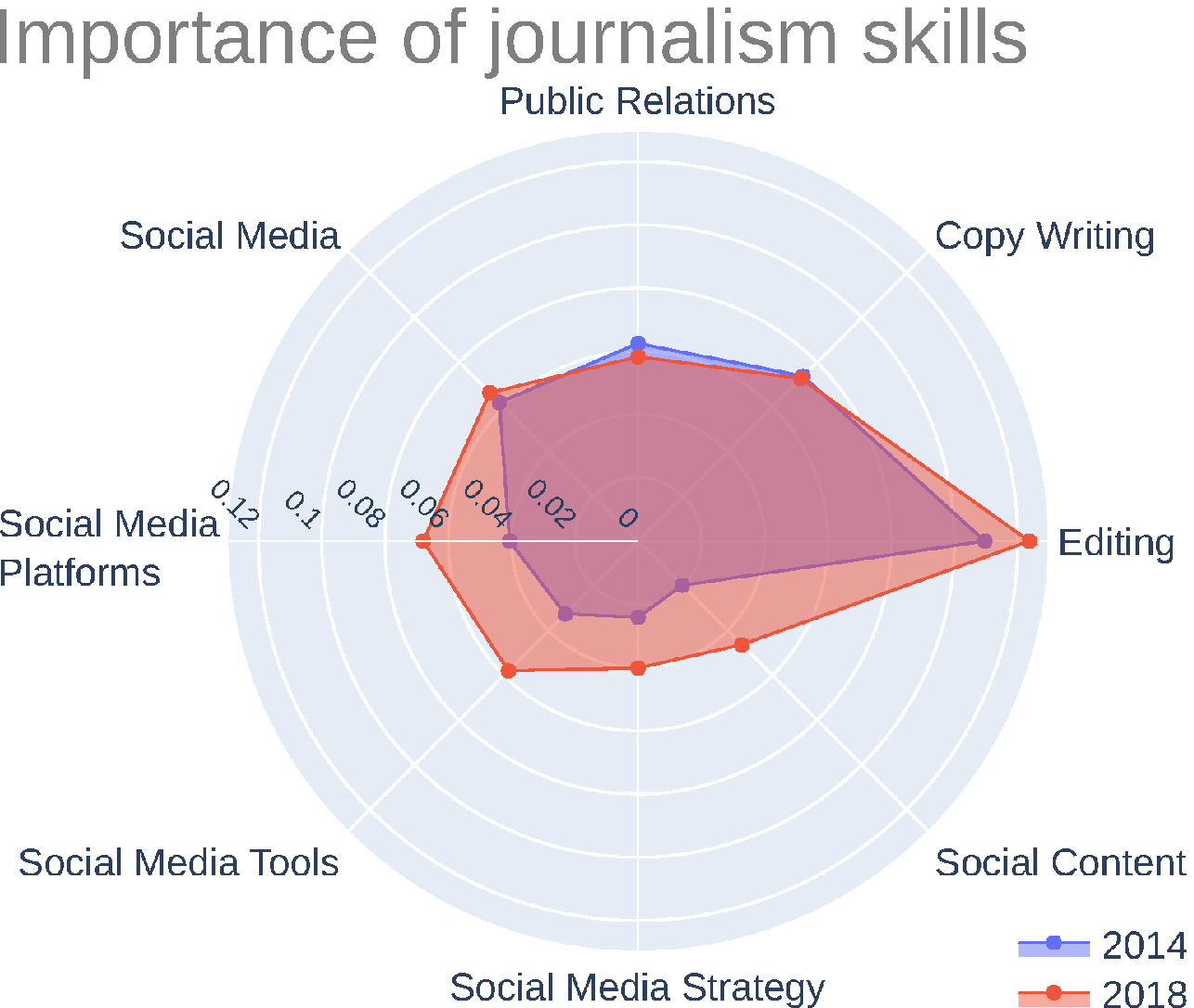}
		\label{subfig:skills-radar}}
	\hspace{0.1cm}
	\subfloat[]{
		\includegraphics[height = \myheight\textheight]{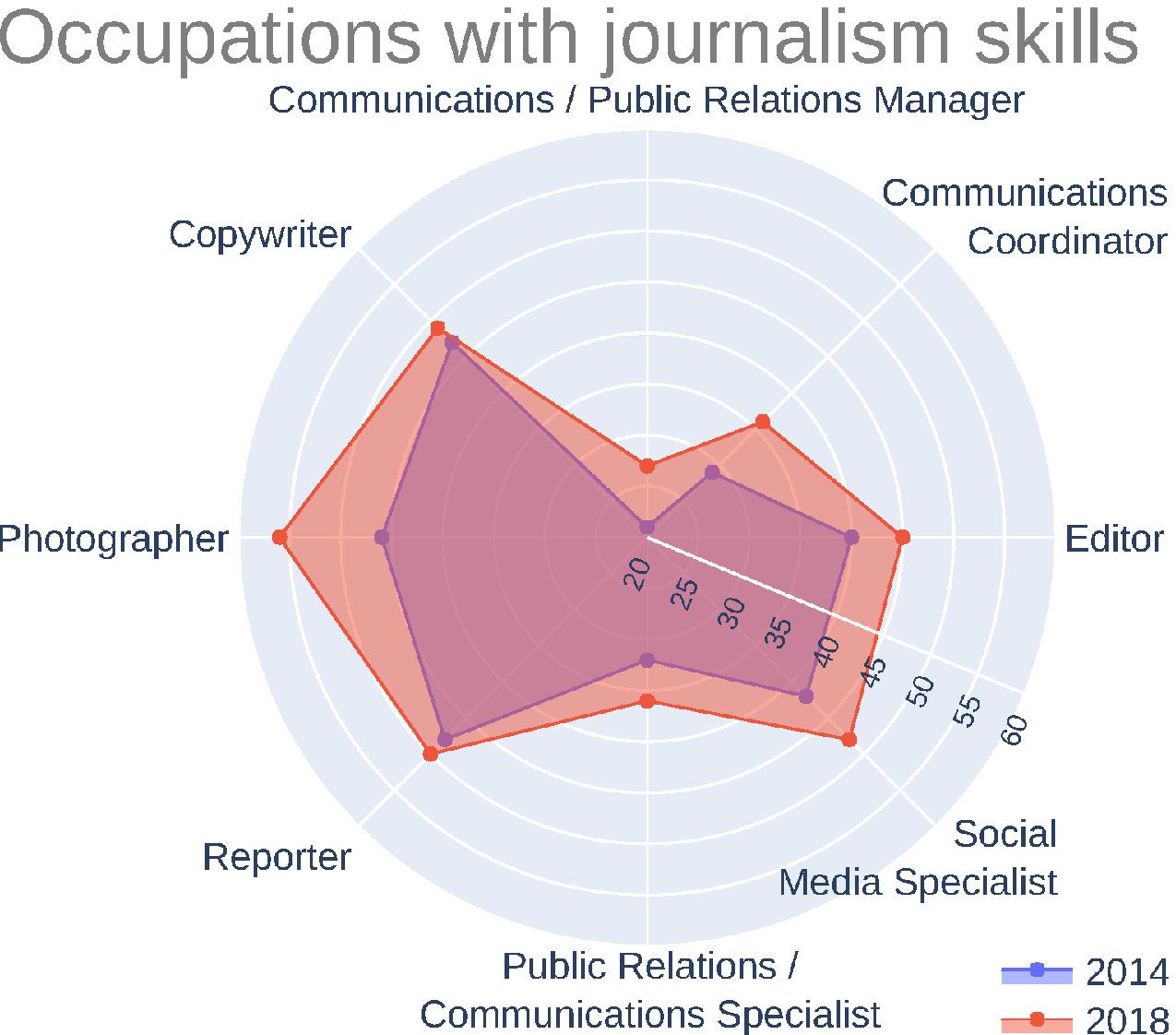}\label{subfig:occupations-radar}
	}\caption{
		\textbf{Skill and occupational similarity analyses}:
	 \textbf{(a)} The changing similarity (or relative importance) of specific skills compared to the skill `Journalism'; 
	 \textbf{(b)} Eight occupations that had the highest similarity to the `Top Yearly Journalism Skills'. 
	 }
	\label{fig:radar}
\end{figure*}

\textbf{Changing importance of journalism skills.} 
We wanted to determine whether the relative importance of the Journalism skill changed over time, using a skill similarity approach.
Given the dynamics of skill requirements in job ads, skills can become increasingly more (or less) similar over time.
We used the similarity measures in \cref{eq:theta} to identify the skills that are becoming more relevant to being a journalist (see ~\textit{\nameref{sec:data-methods}} for details, and the Supplemental Material~\cite{appendix} for the top 50 skills for each from 2014 to 2018). 
The higher the similarity score, the more likely the skills pair will complement and support each other in a given job.
\cref{subfig:skills-radar} shows the changes in similarity scores between the skill `Journalism' and each of the eight other top journalism skills (as per the top yearly journalism skills lists in the Supplemental Material~\cite{appendix}).
The greater the area covered in the radar chart, the greater the similarity score, with the blue area representing 2014 and the red area 2018.\footnote{At the time of analysis, 2018 was the final full year available of access to the required skills-level data.} 
Visibly in \cref{subfig:skills-radar}, `Social Media' related skills became increasingly relevant for journalists, with the relative ratio of more traditional skills such as `Editing' and `Copy Writing' diminishing with respect to `Social Media', from 2014 to 2018.

\textbf{Occupations that require journalism skills.}
Here, we studied which occupations most required journalism skills, and their dynamics over time (according to the BGT occupational taxonomy).
Given the yearly lists of top journalism skills (described in ~\textit{\nameref{subsec:skill-similarity}}), we used \cref{eq:skill-intensity} to determine the occupations with the highest intensities of journalism skills, for each year from 2014 to 2018.
Intuitively, this allows us adaptively to identify occupations that become more or less similar to `Journalism', based on their underlying skill usage. 
It also provides a means to assess likely transitions between occupations, as workers are more likely to transition to occupations where the underlying skill requirements are similar~\citep{Bechichii2018}. 
Higher similarity lowers the barriers to entry from one occupation to another.

\cref{subfig:occupations-radar} highlights eight top occupations and their journalism skill intensity scores for 2014 and 2018. 
`Reporter', `Editor', and `Copywriter' cover the highest percentage of journalism jobs in the dataset, respectively. 
While the journalism skill intensities of these occupations were relatively high in 2018, their growth since 2014 was relatively low. 
In comparison, `Photography', `Communications', `Social Media', and `Public Relations' experienced higher journalism skill intensity growth from 2014 to 2018. 
This provides insights as to where workers with journalism skills might have found employment outside of journalism.

\section{Discussion}
\label{sec:discussion}

\subsection{Volatility of journalism jobs}
\label{subsec:volatility}
Drawn from job ads and employment statistics, our findings reveal the highly volatile nature of the journalism industry. 
Compared to other occupations and the aggregate labour market, journalism experiences dramatic fluctuations that are unpredictable and irregular (see \cref{fig:predictability}). 
The data also confirmed that journalism is an industry in crisis, worsened in the early stages of COVID-19. 
However, the data also reveals surprises, including that the number of journalism jobs ads and employment levels \textit{increased} from 2012 until 2016. Since then, though, journalism jobs in Australia declined.

The volatility of journalism jobs in Australia was clearly apparent in \nameref{subsec:posting-freq-employment}. 
Posting frequency of job ads ranged from near zero levels in 2012 and 2014 to more than 200 posts per quarter in 2016. 
These violent swings are also apparent in the quarterly employment statistics of `Journalists and Other Writers'. 
Following the mass redundancies of 2012, employment levels plummeted, reaching their lowest levels in 2013. They then increased before falling again into the beginning of the COVID-19 pandemic. 
However, the data confirms that volatility of employment has been a constant for journalism, and that this has worsened during COVID-19.

\cref{fig:predictability} reveals this extreme volatility. 
The error metrics from the Machine Learning model used to predict daily posting frequencies of job ads (as detailed in \nameref{subsec:trend-predictability}) highlight the difficulties of making predictions about journalism employment. 
This lack of predictability is indicative of volatility. 
The higher the error scores for a given occupation, the higher the likelihood that the occupation is experiencing significant disruption. 
This becomes apparent when we compare journalism to other occupations. 
For example, the volatility of 'Journalists' dwarfs that of `Data Scientists', an occupation experiencing significant demand and volatility in Australia~\citep{Dawson2019}.

The volatility of journalism jobs was further revealed by a time series analysis of journalism compared to other occupations (\cref{fig:trendlines}), a gender-based analysis (\cref{fig:gender}), a geographical analysis (\cref{fig:location}) and an analysis of the temporary nature of journalism jobs (\cref{fig:temp-vs-perm}).

What is indisputably clear is that the advertising market for news and journalism collapsed, and, at the time of writing, continues to collapse~\citep{shirky2009newspapers, ACCC2019, Doctor_Nieman}.
Meanwhile, consumers have tended to show an unwillingness to pay for digital journalistic content:
in 2019, Australian news consumers admitted they would much would rather subscribe to a video streaming service such as Netflix (34 per cent), than pay for online news (9 per cent)~\citep{DNR2019}. Admittedly, during COVID-19 some subscription rates have risen~\citep{Edmonds2020-nh}. 
Clearly, however, the Internet has detonated the advertising model that once sustained journalism, and simultaneously re-adjusted consumer expectations on the monetary value of journalism content. 
The fact that journalism is struggling is confirmed in several ways by the data, including by the unpredictability of job ads posting frequency and the clear shifts in employment levels, as shown in \cref{fig:quarterly-posts-unit-emp}. 
To say that journalism has been disrupted is an understatement.

\textbf{Volatility exacerbated by COVID-19.}
In a fragmenting news ecosystem, consumer demand for news and journalism is difficult to quantify. 
The \textit{Digital News Report: Australia 2019} has found that many consumers are disengaging, with the proportion of Australians avoiding news increasing from 57 per cent in 2017 to 62 per cent in 2019~\citep{DNR2019}. 
Demand for `quality' and `public interest' journalism is even harder to quantify, given ongoing debates as to what exactly constitutes `quality' and `public interest'~\citep{Wilding2018}. 
Nonetheless, demand for journalism has surged dramatically since the outbreak of COVID-19.

The irony of the coronavirus pandemic is that even as it has been killing off journalism jobs, it has also created a heightened demand for, and appreciation of, journalism among the general public. 
As news analyst ~\citet{Doctor_Nieman} wrote of the US situation in late March, 'The amount of time Americans spend with journalists' work and their willingness to pay for it have both spiked, higher than at any point since Election 2016, maybe before ... [but] how many journalists will still have jobs once the initial virus panic subsides?'. 
In the UK in March, \textit{The Guardian} received 2.17 billion page views, an increase of more than 750 million above its previous record, set in October 2019~\citep{Bedingfield2020}.

Since the outbreak of COVID-19, the volatility of the journalism jobs market has worsened dramatically. 
We noted above that in May News Corp ended the print run of more than 100 newspapers nationally In April, Australian Community Newspapers, which publishes 170 community titles, said it was suspending publication of some of its non-daily newspapers; as a result, four printing presses were closed and an unspecified number of staff were stood down~\citep{Meade2020-sb}. 
Also in April, the federal government announced a AU\$50million package to support public interest journalism across TV, newspapers and radio in regional and remote Australia~\citep{Hayes2020}.
And on April 20, the Australian government announced that digital platforms including Google and Facebook would be forced to pay for content as the Internet advertising business would be overhauled to help local publishers survive the economic fallout of the coronavirus crisis~\citep{Crowe2020}. 
The scheme, which would involve a mandatory code imposed on digital giants, would potentially set a global precedent. 
The combined and ongoing impact on journalism jobs of these sudden, cumulative developments are hard to predict, but will no doubt be profound. 

\subsection{Gender Wage Gap}
\label{subsec:wage-gap} 
At first glance, the data seems to suggest that gender equity is finally arriving in Australia for journalism -- an industry that has traditionally been male-dominated -- as more women than men are employed. 
As the data shows, in 2014 there were 0.7 female journalists employed for every male Journalist, but by 2018 the proportion of female-to-male employment more than doubled, with almost 1.8 female journalists employed for every male Journalist. 
It then declined in 2019 to 1.35, a proportion still almost double that of 2014. 

However, further detail reveals that equity remained elusive. 
Specifically, wage inequality worsened. 
Since 2014, annual salaries for female journalists increased by AU\$3,000, compared with an increase for male journalists of over AU\$30,000 over the same period. 
From 2014 to 2018, average wage growth for Male journalists was more than ten times greater than for female journalists. 
Meanwhile, the average male Journalist was getting older, while the average female Journalist was getting younger. 
In 2014, the average age for a Journalist, whether male or female, was roughly the same: late 30s. 
By 2018, the average age for a male journalist was 42, whereas for a female journalist it was 34.
These results support previous findings on the changing demographic characteristics of journalists in Australia. 
In a survey of female journalists in Australia, \citet{north2016gender} found gendered divisions of tasks associated with reporting, where the majority of female reporters were assigned `soft-news' areas, such as arts, education, and health. 
These gender and age inequities for journalists were also present in other countries~\citep{hanitzsch2019worlds}. 
The wage and age discrepancies between female and male journalists observed in the employment statistics are consistent with the surveyed experiences of female journalists in Australia by~\citet{north2016gender}.

The potential impacts of this worsening disparity are concerning. 
It is possible that senior positions responsible for major editorial decisions were increasingly being dominated by men, whereas junior roles were being filled by women who are younger and worse-paid. 

Further research is needed into related issues of the industry's composition, including, for instance, the ethnicity of journalists. 
A vast body of literature exists regarding the importance of diversity in news~\citep{Rodrigues2018-hn, Budarick2017-qg}. 
Further work is needed into diversity (and its various sub-categories), and what effect diversity has, for instance, on the proportion of people who are actively avoiding the news.

\subsection{Location}
\label{subsec:location-discussion}
As discussed above, the sustained pressures on regional and local journalism have led to a worrying growth of `news deserts' in countries including Australia and the US. 
This trend alarmingly accelerated in the early stages of COVID-19, leaving many areas without any regional or local news coverage. 
For instance, as of October 2020, `Public Interest Journalism Initiative' had documented a net decline of 124 newsrooms from January 2019~\citep{Piji_undated-se}.
Hence, we might assume that journalism jobs in regional and local areas were been drying up, and that an ever-increasing proportion of journalism jobs were in urban centres.

The data, however, were not so clear until the end of 2019. 
As \cref{subfig:cities} shows, in 2012 fewer than a quarter of Australia's journalism job ads were for jobs outside Sydney, Melbourne, Brisbane, Canberra and the ACT or Perth. 
In every subsequent year, the proportion of job ads for journalism positions outside these urban centres was considerably higher. The peak came in 2017, when nearly half of all job ads were for positions outside the major cities. Does this suggest that in 2017 there were as many jobs for journalists in the regions as in the centres? 
Surely not. 
The explanation, we suggest, lies in various factors. 
These include that regional journalism jobs are hard to fill, perhaps because they offer relatively low salaries, and are hence re-advertised. 
It is also possible that there is a high turnover for some regional positions. 
In short, the job ads data may simply be an indication that the journalism industry is even more volatile in the regions than in major urban centres.

Research consistently and emphatically reveals that regional and local journalism have been suffering, with an increasingly bleak prognosis of cuts and closures~\citep{Abernathy2018,ACCC2019,Doctor_Nieman}.
While the data shows a surprisingly high proportion of journalism job ads for positions outside the main metropolitan centres, this cannot be taken to suggest that journalism is holding steady in these areas.

\subsection{Evolving journalism skills}
\label{subsec:evolving-journalism-skills}
Skills are the building blocks of jobs and standardised occupations. 
In this regard, occupations can be characterised as `sets of skills'. 
Intuitively, skills that are similar can be interpreted as complementary when they are paired together or relatively easy to acquire when one skill is already possessed. 

This intuition provides insight into how journalism skills are evolving and where journalists might be finding alternate career paths. 
As \cref{fig:quarterly-posts-unit-emp} shows, both the demand for and supply of journalists have been declining in Australia since 2016. 
Therefore, a growing number of former journalists, who presumably possess an assortment of journalism skills, needed to transition between occupations to find new work. 
There are, however, significant transition costs moving between jobs~\citep{Bechichii2018, Bessen2015}. 
These costs can come in the form of education, training, physically moving for new employment and other barriers. 
To reduce the friction of these transition costs, workers tend to leverage their extant skills, in concert with acquiring new skills, to make career transitions.

As seen in \cref{subfig:skills-radar}, the skill `Journalism' became more similar to `Social Media' and more `generalist' communications skills. 
After applying the \textit{Skill Intensity} formula from \cref{eq:skill-intensity}, we identified the top occupations with highest intensities of journalism skills from 2014-2018 -- these normalised measures from \cref{eq:theta} and \cref{eq:skill-intensity} take into account newly emerging and redundant skills. 
The \cref{subfig:occupations-radar} chart reinforces that top journalism skills were becoming more important to other occupations, such as `Photographers', `Social Media Strategists', `Public Relations Professionals', and `Communications Specialists'.
While it is certainly possible that journalism tasks were being performed in these different occupations, it nonetheless highlights the changing nature of journalism work and the occupations where journalism skills were of growing importance.

From the data, we suggest, three conclusions can be drawn, which supports previous research~\citep{young2018, ORegan2019, macnamara2014journalism, macnamara2016continuing}.
First, to be hired, journalists are required to have a wider array of skills, such as photography and social media aptitude. 
Second, jobs requiring journalism skills were increasingly occupations in social media, generalist communications, and public relations rather than in reporting and editing. 
And third, we see hints as to where onetime journalists are finding alternate career paths. 
As employment conditions progressively worsen, journalists are seemingly pursuing new careers in the occupational areas seen in \cref{subfig:occupations-radar}, such as photography or public relations.

At a time of great uncertainty, with employment prospects deteriorating, it is no wonder that journalists look beyond traditional journalism for their futures. 
For society, however, the implications are significant. 
In this time of economic instability and polarising politics, the people who possess the journalism skills required to keep the public informed and hold leaders to account are, in many cases, employing their talents elsewhere. 
This places enormous strain on the health and quality of journalism in Australia.

\section{Conclusion}

The data reveals a contradiction: demand for journalism skills increased at the same time that demand and employment for journalists declined. 
Indeed, this is one of several contradictions in a volatile industry. 
For an increasing number of news media organisations, a sustainable business model remains elusive.

Our findings give a clearer outline of the problem. Unfortunately, the solutions remain less clear. 
Quality journalism is expensive. 
Good reporting is often slow and laborious, fixed to the unfolding story. 
What is required of quality journalism is, therefore, at odds with the prevailing employment conditions. 
The declines in employment of traditional journalists could have serious implications if the media produce poorer quality materials.

This paper highlights the stresses experienced by journalism in Australia by analysing jobs data. 
We observed the volatility and downward trajectory of the occupation both in job ads and employment statistics. 
These unfavourable employment conditions were worsened by the unfolding COVID-19 crisis. 
Our longitudinal analysis also yields important findings regarding gender inequity. 
While women represented a greater share of employed journalists, they earned less, and the wage gap grew.

Further, this paper also identified top journalism skills. 
Adopting a data-driven method, we described which skills are most similar to `Journalism'. 
We then used these yearly skill sets to adaptively select similar occupations.
This enabled us to quantitatively show that the skill demands of journalists became increasingly similar to those of `Social Media Strategists', `Public Relations Professionals', `Communications Specialists', and others. 
This suggests where people with journalism skills were likely finding alternate career paths, but also raises a related concern. 
On the face of it, the journalism jobs data we have analysed does not look so bad after all. 
On reflection, however, it suggests that the thinning ranks of `journalism' are populated by fewer journalists, and more public relations specialists.

Future research could compare these results to other labour markets in different countries to assess and compare the validity of these findings. 
For example, the skill similarity methodology could be applied in other labour markets to compare the resulting top journalism skills in different locations. 
Additionally, labour demand analyses could be conducted on occupations most similar to journalists to better understand the incentives to transition to other vocations. This could provide insights into the boundaries of `journalism work' and analyse the relative demand for different types of journalism.
Further work could also examine the implications of changing journalism skill demands for journalism schools. This research demonstrated that not only have the skills demanded of journalists evolved, but the occupations that require journalism skills have broadened. The extent to which journalism schools are adequately preparing their students for the quickly changing labour demands of journalists is a rich area of inquiry.

The results from this research both reinforce the well-documented difficulties of journalism in Australia and provide granular details that isolate and reveal these challenges. The hope is that these analytical methods and insights can contribute to the health and well-being of the Fourth Estate, and hence to the health and well-being of society.

\begin{acks}
	We would like to thank Burning Glass Technologies for generously providing the job advertisements data that has enabled this research.
\end{acks}

\begin{funding}
	The authors disclosed receipt of the following financial support for the research, authorship, and/or publication of this article: Marian-Andrei Rizoiu was partially supported by Facebook Research under the Content Policy Research Initiative grants and the Commonwealth of Australia (represented by the Defence Science and Technology Group).
\end{funding}

\begin{biogs}
	\textbf{Nikolas Dawson} is a Labour Economist and PhD candidate at the University of Technology Sydney. His doctoral research at UTS is on the changing labour market dynamics in Australia, where he analyses issues such as skill shortages, job transitions, and emerging technology adoption in the Australian labour market. In his research, Nikolas applies data science and machine learning techniques to draw insights from large datasets, including online job advertisements, employment statistics, and longitudinal household survey data. During his Doctoral studies, Nikolas was selected as an OECD Future of Work Research Fellow and has also worked with agencies of the United Nations in Geneva, Switzerland researching the `Economic Impacts of Artificial Intelligence'.
	
	\noindent\textbf{Sacha Molitorisz} is a lecturer at the Centre for Media Transition with the University of Technology Sydney, employed jointly by the Law Faculty and the Faculty of Arts and Social Sciences. In May 2020, his book Net Privacy, How we can be free in an age of surveillance, was published by New South Books (Australia) and McGill-Queens University Press (Canada). He has co-authored com-missioned reports for bodies including the Australian Competition and Consumer Commission and the Australian Communications and Media Authority, he has worked as a consultant for the Public Interest Journalism Initiative on the development of Australia’s world-first news media bargaining code, and he has been published in mainstream outlets such as The Conversation and The Sydney Morning Herald as well as in leading academic journals. He is also a regular media commentator. Previously, he spent nearly 20 years on staff at the Sydney Morning Herald and smh.com.au as a writer, editor, and blogger. 
	
	\noindent\textbf{Marian-Andrei Rizoiu} is a lecturer with the University of Technology Sydney, where he leads the Behavioral Data Science group, studying human attention dynamics in the online environment. His research has made several key contributions to applications such as online popularity prediction and real-time tracking and countering disinformation campaigns. For the past four years, he has been developing theoretical models for online information diffusion, which can account for complex social phenomena, such as the rise and fall of online popularity, the spread of misinformation, or the adoption of disruptive technologies. He approached questions such as ``Why did X become popular, but not Y?'' and ``How can problematic content be detected based solely on how it spreads?'' with implications in detecting the spread of conspiracy theories and disinformation campaigns. Marian-Andrei also works in understanding shortages and mismatches in labour mar-kets. Together with his collaborators, he built a real-time occupation transition recommender system usable in periods of massive disruptions (such as COVID-19). He also has shown that social-media predicted personality profiles are linked with worker occupation. Marian-Andrei has published in the most selective venues, such as the PNAS, WWW, WSDM, ICWSM, and CIKM. His work has received significant media attention---including Bloomberg Business Week, Nature Index, BBC, and The Conversation---, and Marian-Andrei served as an expert for the NSW government's Defamation Law Reform. See more at \url{http://www.rizoiu.eu}.
	
	\noindent\textbf{Peter Fray} is managing editor of Private Media, an independent Australian news media company. He was formerly a professor of journalism practice at University Technology Sydney, the founder of PolitiFact Australia, and the editor of The Sydney Morning Herald, the Canberra Times, and the Sunday Age.
\end{biogs}

\bibliographystyle{SageH}

\clearpage
\appendix
\etocdepthtag.toc{mtappendix}
\etocsettagdepth{mtchapter}{none}
\etocsettagdepth{mtappendix}{subsection}
\etoctocstyle{1}{Contents (Appendix)}
\tableofcontents

This document is accompanying the submission \textit{\titlename}.
The information in this document complements the submission, and it is presented here for completeness reasons.
It is not required for understanding the main paper, nor for reproducing the results.

\section{Technical Appendix}
\label{sec:appendix1}
Here, we describe the data sources we used to analyse journalism jobs. We also outline the skill similarity methodology that enables us to construct temporal (yearly) sets of top journalism skills. Lastly, we describe how these temporal sets of top journalism skills then allow us to adaptively identify occupations that are `most similar' to journalism, at the granular skill level.

\subsection{Data Sources}
\label{subsec:appendix-data}

\textbf{Journalism job ads.} This research draws on more than 8 million Australian online job ads from 2012-01-01 until 2019-02-28, courtesy of data provided by Burning Glass Technologies\footnote{BGT is a leading vendor of online job ads data. \texttt{https://www.burning-glass.com/}} (BGT). 
BGT also granted access to the aggregated job ads data from 2019-03-01 to 2020-03-31, allowing us to address the early impacts of the unfolding coronavirus pandemic (COVID-19) on journalism jobs in Australia. 
BGT collected the job ads data via web scraping and systematically processed it into structured formats. The dataset consists of detailed information on individual job ads, such as location, salary, employer, educational requirements, experience demands, and more. The skill requirements have also been extracted (totalling $>11,000$ unique skills) and each job ad is classified into its relevant occupational and industry classes.
There are two occupational ontologies in the job ads dataset. The first is ANZSCO, which is the official occupational classification standard in Australia and New Zealand. The other is the BGT occupational ontology, which has been developed due to shortcomings of official occupational standards (as described in \nameref{sec:related-work}). 

To ensure selection accuracy, we instituted the following search query conditions over the dataset:

\begin{enumerate}
    \item All job ads with ANZSCO Occupation labels of `Newspaper or Periodical Editor', `Print Journalist', `Radio Journalist', `Television Journalist', and `Journalists and Other Writers nec'
(where `nec' stands for `not elsewhere classified').
\item OR All job ads with the BGT Occupation label of `Journalist / Reporter' and `Editor' (the two primary BGT occupational classes for journalists);
    \item OR All job ads with the `Journalist', `News', or `Editor' in any part of the job title.
\end{enumerate}

After manually reviewing the returned job ad features for accuracy, the selection process resulted in a sample of 3,231 Australian journalism job ads from 2012-01-01 until 2019-02-28. 
We used the same search query and approach for the 2019-03-01 to 2020-03-31 period to supplement this sample. This returned 467 journalism job ads, amounting to a total of 3,698 journalism job ads from 2012-01-01 to 2020-03-31.
The job ads during the period are observed aggregated daily, with limited skill level details.
However, much of the analysis that follows requires access to the features within individual job ads, so only \cref{fig:covid-posting-freq} leverages the 2020 data.

\textbf{Further details on job ads data.}
It is estimated that approximately 60\% of Australian job ads are posted online~\citep{Department_of_Employment_Skills_Small_and_Family_Business_undated-qv}, which grew quickly in the early 2000's before plateauing in recent years~\citep{Carnevale2014-xc}. At aggregate levels, online job advertisements (ads) provide valuable indicators of relative labour demands. This includes demand features, such as salaries, educational requirements, years of experience, and, most importantly, skill-level information. Here, a distinction must be made between skills, knowledge, abilities, and occupations. `Skills' are the proficiencies developed through training and/or experience~\citep{Oecd2019-cl}; `knowledge' is the theoretical and/or practical understanding of an area; `ability' is the competency to achieve a task~\citep{Gardiner2018-dt}; and `occupations' are standardised jobs that are the amalgamation of skills, knowledge, and abilities used by an individual to perform a set of tasks that are required by their vocation. Throughout this paper, the term `skill' will incorporate `knowledge' and `ability'. Skills, in this sense, are the constituent elements that workers use to perform tasks, which ultimately define jobs and occupations.
While it is possible that the ways skills are described could evolve over time, it is unlikely that their meanings materially changed over the nine year period analysed in this research.

\textbf{Advantages of job ads data.} Understanding how the composition of skill sets evolve within an occupation is essential to understanding trends in that occupation. However, occupational data rarely captures skill-level data. Most often, official occupational standards are static, rarely updated classifications, which fail to capture the changing skill demands of occupations, or to detect the creation of new types of jobs. 

\textbf{Example of journalism job ad titles.} The table below illustrates a random sample of job ad titles classified as journalists in the BGT dataset.

\begin{table}
	\centering
	\caption{Random sample of journalism job ad titles}
	\begin{tabular}{p{0.45\textwidth}}
		\toprule
		Job Ad Title \\
		\midrule
		News And Features Journalist - The Courier \\
		Journalist/Copywriter \\
		Senior Finance Journalist/Content Manager \\
		ABC Rural Reporter \\
		Rural Reporter, ABC Local Radio \\
		Newspaper Journalist \\
		Journalist Radio News - Illawarra \\
		News Writer \\
		Editor, News \\
		Senior Digital News Editor - Leading Digital Media Company \\
		\bottomrule
	\end{tabular}
\end{table}

\textbf{Journalist employment statistics.} Employment data (labour supply) were collected from the `Quarterly Detailed Labour Force' statistics by the ABS~\citep{Australian_Bureau_of_Statistics2019-sv}. These employment data are organised into standardised occupations called the Australia and New Zealand Standard Classification of Occupations (ANZSCO). ANZSCO provides a basis for the standardised collection, analysis and dissemination of occupational data for Australia and New Zealand. The structure of ANZSCO has five hierarchical levels - major group, sub-major group, minor group, unit group and occupation.  The categories at the most detailed level of the classification are termed 'occupations'.

A shortcoming, however, is that the lowest level of occupational employment data available by the ABS is at the 4-digit Unit level, which is one hierarchical level above specific occupations. As our research is focused on the employment Unit class of `Journalists and Other Writers', all ABS employment statistics cited in this research include the following occupations: `Copywriter', `Newspaper or Periodical Editor', `Print Journalist', `Radio Journalist', `Technical Writer', `Television Journalist', and `Journalists and Other Writers nec'. 
While the inclusion of the `Copywriter' and `Technical Writer' occupations in these statistics could distort results pertaining to `Journalists' to an extent, we consider this impact to be limited in scope.
As we describe in \textit{\nameref{sec:jobs-data-analysis}}, the employment statistics highlight important trends in journalism occupations, which are confirmed by findings from the job ads data.

Another shortcoming of employment statistics is their `lagging' nature. The inertia of labour markets means that it takes time for changes to materialise in employment statistics. Additionally, the official reporting of employment statistics takes time. 
Employment statistics are often published several months or years after the reported period. 
As a result, these `lagging' characteristics are not available for the most recent periods in our work (such as for the second half of 2019 and later.)

\subsection{Skill Similarity}
\label{subsec:appendix-skill-similarity}

In this section, we detail the methodology previously employed in \citep{Alabdulkareem2018-jl, Dawson2019} to measure skill similarity dynamically.
Here, we present the building blocks for this method, applying it for journalism related skills and occupations.

\textbf{Intuition.}
Two skills are similar when the two are related and complementary, i.e. the two skills in a skills-pair support each other. 
For example, `Journalism' and `Editing' have a high pairwise similarity score because together they enable higher productivity for the worker, and because the difficulty to acquire either skill when one is already possessed by a worker is relatively low.

Our goal, therefore, is to calculate the similarity of each unique skill relative to every other unique skill in the dataset. Such a measure allows us to identify which skills have the highest pairwise similarities to a specific skill or set of skills. We also want to identify how skill similarity evolves over time. To achieve this, we have instituted a temporal split of a calendar year. This enables us to assess yearly changes to the underlying skill demands of journalism jobs.

\textbf{The Revealed Comparative Advantage of a skill.}
We implement a data-driven methodology to measure the pairwise similarity between pairs of skills that co-occur in job ads. 
One difficulty we encounter is that some skills are ubiquitous, occurring across many job ads and occupations. 
We address this issue by using the \emph{Revealed Comparative Advantage} (RCA), which
maximises the amount of skill-level information obtained from each job ad, while minimising the biases introduced by over-expressed skills in job ads.
Formally, RCA measures the relevance of a skill $s$ for a particular job ad $j$ as:
\begin{equation} \label{eq:rca}
  RCA(j, s) = \frac{x(j, s) / \mathop{\sum}\limits_{s'\in \mathcal{S}}x(j, s')}
  {\mathop{\sum}\limits_{j'\in J}x(j', s) / \mathop{\sum}\limits_{j'\in \mathcal{J},s'\in \mathcal{S}}x(j', s')}
\end{equation}
\noindent
where $x(j,s) = 1$ when the skill $s$ is required for job $j$, and $x(j,s) = 0$ otherwise;
$\mathcal{S}$ is the set of all distinct skills, and $\mathcal{J}$ is the set of all job ads in our dataset.
$RCA(j, s) \in \left[ 0, \mathop{\sum}\limits_{j'\in J,s'\in S}x(j', s') \right], \forall j, s$, and the higher $RCA(j, s)$ the higher is the comparative advantage that $s$ is considered to have for $j$.
Visibly, $RCA(j, s)$ decreases when the skill $s$ is more ubiquitous (i.e. when $\mathop{\sum}\limits_{j'\in J}x(j', s) $ increases), or when many other skills are required for the job $j$ (i.e. when $\mathop{\sum}\limits_{s'\in S}x(j, s')$ increases).
$RCA$ provides a method to measure the importance of a skill in a job ad, relative to the total share of demand for that skill in all job ads. It has been applied across a range of disciplines, such as trade economics~\citep{Hidalgo2007-qk}~\citep{Vollrath1991-kr}, identifying key industries in nations~\citep{Shutters2016-fe}, and detecting the labour polarisation of workplace skills~\citep{Alabdulkareem2018-jl}.

\textbf{Measure skill similarity.}
The next step is measuring the complementarity of skill-pairs that co-occur in job ads. 
First, we compute the `effective use of skills'
$e(j, s)$ defined as $e(j, s) = 1 \text{ when } RCA(j,s) > 1$ and $e(j, s) = 0$ otherwise.
Finally, we compute the skill complementarity (denoted $\theta$) as the minimum of the conditional probabilities of a skills-pair being effectively used within the same job ad. 
Skills $s$ and $s'$ are considered as highly complementary if they tend to commonly co-occur within individual job ads, for whatever reason. Formally:
\begin{equation} \label{eq:theta-si}
  \theta(s, s') = \frac{\mathop{\sum}\limits_{j'\in J}e(j,s).e(j,s')}
  {max \left( \mathop{\sum}\limits_{j'\in J}e(j,s), \mathop{\sum}\limits_{j'\in J}e(j,s') \right)}
\end{equation}
Note that $\theta(s, s') \in [0, 1]$, a larger value indicates that $s$ and $s'$ are more similar, and it reaches the maximum value when $s$ and $s'$ always co-occur (i.e. they never appear separately).

\textbf{Top journalism skills.} 
Following the procedure outlined in \citep{Dawson2019} for building sets of highly complementary skills,
we use the $\theta$ function together with `Journalism' as the `seed' skill to create top yearly lists of journalism skills. 
More precisely, we compute $\theta(Journalism, s)$ -- i.e. the similarity between the skill `Journalism' and each unique skill that occurs during a given year.
Skills on each yearly list are
ordered by their descending pairwise skill similarity scores. 
When inspecting the yearly skill lists, we make two observations. 
First, the skills in 2012 and 2013 appear of notably lower quality than from 2014 onward. 
We posit that this has to do with imperfect skills extraction methods during the early years of the BGT dataset.
As a result, we decided to measure the top yearly journalism skill sets from 2014 to 2018 (the last available full year of data for which we had access).~\footnote{We did not notice a deterioration of quality regarding other features, such as salaries, education, experience etc. Therefore, these 2012 and 2013 will be used for parts of the analysis.}
Second, we decided to retain only the top 50 skills on each yearly list.
Through qualitative analysis, we determined that this threshold of 50 is
both sufficiently exclusive for defining journalism skills and reasonably inclusive for detecting the evolution of new, emerging skills in journalism. 
The purpose of these top journalism skills lists is to capture journalism labour trends; it is not intended to represent a complete taxonomy of journalism skills. The yearly lists of top journalism skills, and their similarity scores, can be observed in the Supplemental Material Sec.~\textit{\nameref{sec:appendix2}}.

\textbf{Compute journalism skill intensity.}
For the occupational similarity analysis in Sec.~\textit{\nameref{subsec:journalism-skills}}, we decided to use the BGT occupational ontology as opposed to ANZSCO. 
This is because the BGT occupational classes appear more reflective of current job titles. For example, a job title advertised for a `Social Media Manager' is classified by BGT as a `Social Media Strategist / Specialist'. Whereas the same job title would be classified by ANZSCO as an `Advertising Specialist' or `Marketing Specialist'. 

\subsection{Trend Analysis \& Predictability}
\label{subsec:appendix-trend-analysis-technical}
We use the Prophet time-series forecasting tool developed by Facebook Research~\citep{taylor2018forecasting}. 
Prophet is an auto-regressive tool that fits non-linear time-series trends with the effects from daily, weekly, and yearly seasonality, and also holidays. 
The main model components are represented in the following equation:
\begin{equation} \label{eq:prophet}
    y(t) = g(t) + s(t) + h(t) + \epsilon_t    
\end{equation}
\noindent 
where $g(t)$ refers to the trend function that models non-periodic changes over time; $s(t)$ represents periodic changes, such as seasonality; $h(t)$ denotes holiday effects; and $\epsilon_t$ is the error term and represents all other idiosyncratic changes. 

\subsection{Quantify Labour Demand Volatility}
\label{subsec:appendix-labour-demand-volatility-technical}
We evaluate the forecasting performance using a temporal holdout setup. That is, we split the available time-series into a training part (the first part of the sequence) and a testing part (the latter part of the sequence).
We train the Prophet model on the training part, and we generate job ad posting forecasts by ``running time forward'' in \cref{eq:prophet} for time $t$ in the testing period.
Finally, we measure the accuracy of the forecast against the observed posting volumes using the Symmetric Mean Absolute Percentage Error (SMAPE)~\citep{Scott_Armstrong1985-tt, makridakis1993accuracy}. 
SMAPE is formally defined as:
\begin{equation}
    SMAPE(A_t,F_t) = \frac{200}{T} \sum_{t=1}^{T} \frac{|F_t - A_t|}{(|A_t|+|F_t|)}
\end{equation}
\noindent
where $A_t$ denotes the actual value of jobs posted on day $t$, and $F_t$ is the predicted value of job ads on day $t$. SMAPE ranges from 0 to 200, with 0 indicating a perfect prediction and 200 the largest possible error. When actual and predicted values are both 0, we define SMAPE to be 0. 
We selected SMAPE as an alternative to the more widely used MAPE because it is (1) scale-independent and (2) robust to actual or predicted zero values.
To evaluate the uncertainty of the forecast, we adopt a `sliding window' approach. 
This consists of using a constant number of training days (here $1,186$ days) to train the model, and we test the forecasting performance on the next $365$ days. 
We then shift both the training and the testing periods right by one day, and the process is repeated. Consequently, we train and test the model 365 times, and we obtain 365 SMAPE performance values.

\subsection{Top Journalism Skills by Year}
\label{sec:appendix2}

Top journalism skills calculated by skill similarity methodology in Sec. \textit{\nameref{subsec:skill-similarity}}.
\vspace{.5cm}

\begin{table*}[htbp]
	\centering
	\caption{Top journalism skills calculated by skill similarity methodology in \nameref{subsec:skill-similarity}}
	\begin{adjustbox}{width=1\textwidth}
\begin{tabular}{rlrrrrr}
  \toprule
Rank & 2014                             & 2015                             & 2016                             & 2017                             & 2018                            \\ 
  \midrule
1    & Journalism                       & Journalism                       & Journalism                       & Journalism                       & Journalism                       \\
2    & Editing                          & Editing                          & Editing                          & Editing                          & Editing                          \\
3    & Media Relations                  & Media Relations                  & Copy Writing                     & Content Management               & Content Management               \\
4    & Corporate Communications         & Copy Writing                     & Media Relations                  & Social Media                     & Media Relations                  \\
5    & Copy Writing                     & Content Management               & Content Management               & Copy Writing                     & Copy Writing                     \\
6    & Content Management               & Copywriting                      & Social Media                     & Media Relations                  & Social Media Platforms           \\
7    & Public Relations                 & Social Media                     & Social Media Platforms           & Corporate Communications         & Social Media                     \\
8    & Social Media                     & Public Relations                 & Copywriting                      & Social Media Platforms           & Content Development              \\
9    & Content Management Systems (CMS) & Social Media Platforms           & Corporate Communications         & Content Development              & Corporate Communications         \\
10   & Multimedia                       & Corporate Communications         & Public Relations                 & Social Content                   & Public Relations                 \\
11   & Copywriting                      & Content Development              & Content Development              & Public Relations                 & Social Media Tools               \\
12   & Content Development              & Content Management Systems (CMS) & Social Media Tools               & Copywriting                      & Copywriting                      \\
13   & Strategic Communications         & Strategic Communications         & Digital Marketing                & Facebook                         & Content Management Systems (CMS) \\
14   & Facebook                         & Social Media Tools               & Online Marketing                 & Strategic Communications         & Social Content                   \\
15   & Social Media Platforms           & Multimedia                       & Multimedia                       & Social Media Tools               & Strategic Communications         \\
16   & Marketing Communications         & Facebook                         & Strategic Communications         & Marketing Communications         & Social Media Strategy            \\
17   & Media Coverage                   & Marketing Communications         & Market Research                  & Content Management Systems (CMS) & Content Marketing                \\
18   & Publicity                        & Social Content                   & Marketing Communications         & Multimedia                       & Facebook                         \\
19   & Proofreading                     & Digital Communications           & Content Management Systems (CMS) & Proofreading                     & Digital Communications           \\
20   & Social Media Tools               & Publicity                        & Writing                          & Content Marketing                & Media Coverage                   \\
21   & Digital Communications           & Media Production                 & Content Marketing                & Digital Journalism               & Publicity                        \\
22   & Crisis Management                & Social Media Strategy            & Photography                      & Digital Communications           & Proofreading                     \\
23   & Adobe Photoshop                  & Communications Programmes        & Instagram                        & Publicity                        & Multimedia                       \\
24   & Communications Programmes        & Media Coverage                   & Publicity                        & Media Coverage                   & Instagram                        \\
25   & Digital Journalism               & Internal Communications          & Digital Communications           & Digital Marketing                & Video Production                 \\
26   & Community Relations              & Content Marketing                & Media Coverage                   & Writing                          & Marketing Communications         \\
27   & Photography                      & Proofreading                     & Social Content                   & Video Production                 & Adobe Photoshop                  \\
28   & Social Media Strategy            & Writing                          & Social Media Strategy            & Graphic Design                   & Content Curation                 \\
29   & Graphic Design                   & Adobe Photoshop                  & Media Production                 & Media Production                 & Video Editing                    \\
30   & Youtube                          & Brand Awareness Generation       & Proofreading                     & Communications Programmes        & Adobe Indesign                   \\
31   & Media Strategy                   & Adobe Indesign                   & Facebook                         & Instagram                        & Adobe Creative Suite             \\
32   & Brand Management                 & Marketing Materials              & Event Planning                   & Social Media Strategy            & Adobe Acrobat                    \\
33   & Web Content Management           & Digital Marketing                & Adobe Photoshop                  & Video Editing                    & Brand Awareness Generation       \\
34   & Adobe Indesign                   & Video Editing                    & Meeting Deadlines                & Adobe Photoshop                  & Adobe Illustrator                \\
35   & Social Content                   & Adobe Creative Suite             & Self-Starter                     & Self-Starter                     & Google Analytics                 \\
36   & Marketing Materials              & Adobe Acrobat                    & Marketing                        & Breaking News Coverage           & Press Releases                   \\
37   & Event Planning                   & Graphic Design                   & Creativity                       & Creativity                       & LinkedIn                         \\
38   & Digital Marketing                & Video Production                 & Adobe Indesign                   & Adobe Illustrator                & Digital Marketing                \\
39   & Writing                          & Instagram                        & Adobe Creative Suite             & Event Planning                   & Digital Journalism               \\
40   & Instagram                        & LinkedIn                         & Adobe Illustrator                & Adobe Indesign                   & Media Production                 \\
41   & Online Research                  & Media Strategy                   & Community Relations              & Adobe Creative Suite             & Communications Programmes        \\
42   & Adobe Acrobat                    & Photography                      & Adobe Acrobat                    & Adobe Acrobat                    & Crisis Management                \\
43   & LinkedIn                         & PR Agency                        & Press Releases                   & Promotional Materials            & Media Strategy                   \\
44   & Google Analytics                 & Meeting Deadlines                & Internal Communications          & Photography                      & Writing                          \\
45   & Video Editing                    & Digital Journalism               & Campaign Management              & Content Curation                 & Photography                      \\
46   & Website Production               & Event Planning                   & Creative Writing                 & Marketing                        & Blog Posts                       \\
47   & Proofing                         & Google Analytics                 & Video Production                 & Meeting Deadlines                & Internal Communications          \\
48   & Video Production                 & Media Campaigning                & Blog Posts                       & Google Analytics                 & Event Planning                   \\
49   & Media Planning                   & Press Releases                   & Crisis Management                & Media Strategy                   & Creative Problem Solving         \\
50   & Campaign Management              & Crisis Management                & Youtube                          & Business-to-Business             & Creativity                       \\
   \bottomrule
\end{tabular}
\end{adjustbox}
\end{table*}
 \end{document}